\documentclass[onecolumn,10pt,twoside]{svjour3}

\usepackage{subcaption}
\usepackage{hyperref}
\usepackage{graphicx}
\usepackage{listings}
\usepackage{xcolor}
\usepackage{float}

\def\makeheadbox{\relax}

\date{}

\begin{document}

\title{MementoEmbed and Raintale for Web Archive Storytelling}

\author{Shawn M. Jones \and Martin Klein \and 
Michele C. Weigle \and Michael L. Nelson}
\institute{
  Shawn M. Jones \and Martin Klein \at Los Alamos National Laboratory, Los Alamos, NM
  \and
  Michele C. Weigle \and Michael L. Nelson \at Old Dominion University, Norfolk, VA
}

\maketitle

\begin{abstract}
For traditional library collections, archivists can select a representative sample from a collection and display it in a featured physical or digital library space. Web archive collections may consist of thousands of archived pages, or mementos. How should an archivist display this sample to drive visitors to their collection? Search engines and social media platforms often represent web pages as cards consisting of text snippets, titles, and images. Web storytelling is a popular method for grouping these cards in order to summarize a topic. Unfortunately, social media platforms are not archive-aware and fail to consistently create a good experience for mementos. They also allow no UI alterations for their cards. Thus, we created MementoEmbed to generate cards for individual mementos and Raintale for creating entire stories that archivists can export to a variety of formats.
\end{abstract}

\section{Introduction}
Trying to understand the differences between web archive collections can be onerous. Thousands of collections exist \cite{jones_many_2018}, collections can contain thousands of documents, and many collections contain little metadata to assist the user in understanding their contents \cite{jones_social_2019}. How can an archivist display a sample of a collection to drive visitors to their collection or provide insight into their archived pages?

Search engines and social media platforms have settled on the card visualization paradigm, making it familiar to most users. Web storytelling is a popular method for grouping these cards to summarize a topic, as demonstrated by tools such as Storify. For this reason, AlNoamany et al. \cite{alnoamany_generating_2017} made Storify the visualization target of their web archive collection summaries. Because Storify shut down in 2018 \cite{jones_storify_2017}, we evaluated 62 alternative tools, such as Facebook, Pinboard, Instagram, Sutori, and Paper.li. We found that they are not reliable for producing cards from archived web pages, mementos. Thus we developed MementoEmbed \cite{jones_preview_2018}, an archive-aware service that can generate different surrogates \cite{capra_augmenting_2013} for a given memento.  Currently supported surrogates include social cards, browser thumbnails (screenshots) \cite{kopetzky_visual_1999}, word clouds, and animated GIFs of the top-ranked images and sentences. MementoEmbed's cards appropriately attribute content to a memento's original resource separately from the archive, including both the original domain and its favicon from the memento's time period, as well as providing a striking image, a text snippet, and a title. MementoEmbed provides an extensive API that helps machine clients request specific information about a memento. Raintale \cite{jones_raintale_2019} leverages this API to generate complete stories containing the surrogates of many different mementos.

We review why existing platforms are not reliable for storytelling with mementos, and then detail how MementoEmbed and Raintale work to solve this functionality gap.

\section{Background}

MementoEmbed and Raintale focus on content from web archives. Web archives attempt to capture and preserve \textbf{original resources} from the web, such as web pages and their embedded images, stylesheets, and JavaScript. Each capture comes from a specific moment in time and, as such, is a version of that original resource, a \textbf{memento}. Its capture datetime is its \textbf{memento-datetime}. Lists of mementos for a specific original resource are available as \textbf{TimeMaps}. For clarity in this document, the URIs that identify original resources are  URI-Rs, those that identify mementos are URI-Ms, and those that identify TimeMaps are URI-Ts.  The \textbf{Memento Protocol} \cite{van_de_sompel_rfc_2013} documents all of these concepts. The Memento Protocol also details the \textbf{datetime negotiation} process where a client can supply their desired datetime and URI-R to a \textbf{TimeGate} and receive a matching URI-M closest to that datetime.

Web archives often visualize their holdings for end users via a playback engine that attempts to reconstruct the web page and its embedded resources as closely as possible to their state at the time of capture. This playback engine often applies transformations to the page to make this possible, rewriting links to the mementos of images captured by the web archive instead of images on the live web. The engine also applies branding to each memento in the form of a banner. For analysis, however, we often want to process \textbf{raw mementos} \cite{jones_raw_2016,jones_raw_2016_2,raw_mementos_2017} that do not contain these augmentations.

\begin{figure}[t]
  \centering
  \includegraphics[width=0.8\textwidth]{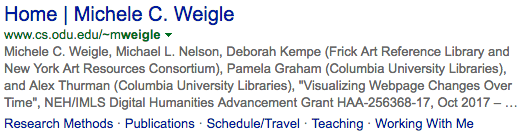}
  \caption{An example surrogate provided by the Bing search engine.}
  \label{fig:bing-text-snippet}
% \end{figure}

% \begin{figure}[t]
%   \centering
  \includegraphics[width=0.8\textwidth]{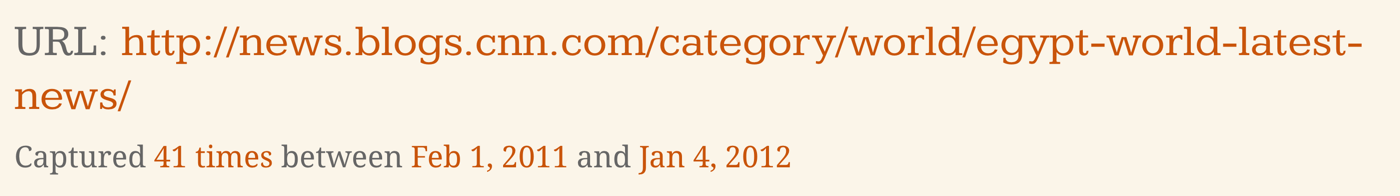}
  \caption{The surrogate used for Archive-It mementos provides the URI-R and capture dates. Other information is optional and not drawn from the underlying page, but supplied by archivists, making them each a poor, inconsistent summary of a memento.}
  \label{fig:archiveitlike-example}
\end{figure}

Once these augmentations are removed, before we can perform any analysis, we still need to remove the HTML code and navigational elements, the \textbf{boilerplate} \cite{pomikalek_removing_2011,kohlschutter_boilerplate_2010,nwala_survey_2017}, from the web page content. With this boilerplate removed we can generate a \textbf{surrogate} \cite{capra_augmenting_2013,al_maqbali_evaluating_2010}, a small summary of the underlying document. Figure \ref{fig:bing-text-snippet} displays a surrogate produced by Bing for a live web resource, and Figure \ref{fig:archiveitlike-example} displays a surrogate produced for a memento from Archive-It. As shown in the Archive-It surrogate, there is little information about the resource. More than 50\% of all Archive-It surrogates resemble this one \cite{jones_social_2019}, making Archive-It's existing surrogates poor summaries of their underlying mementos.

\begin{figure}[htbp]
\centering
\includegraphics[width=0.6\textwidth]{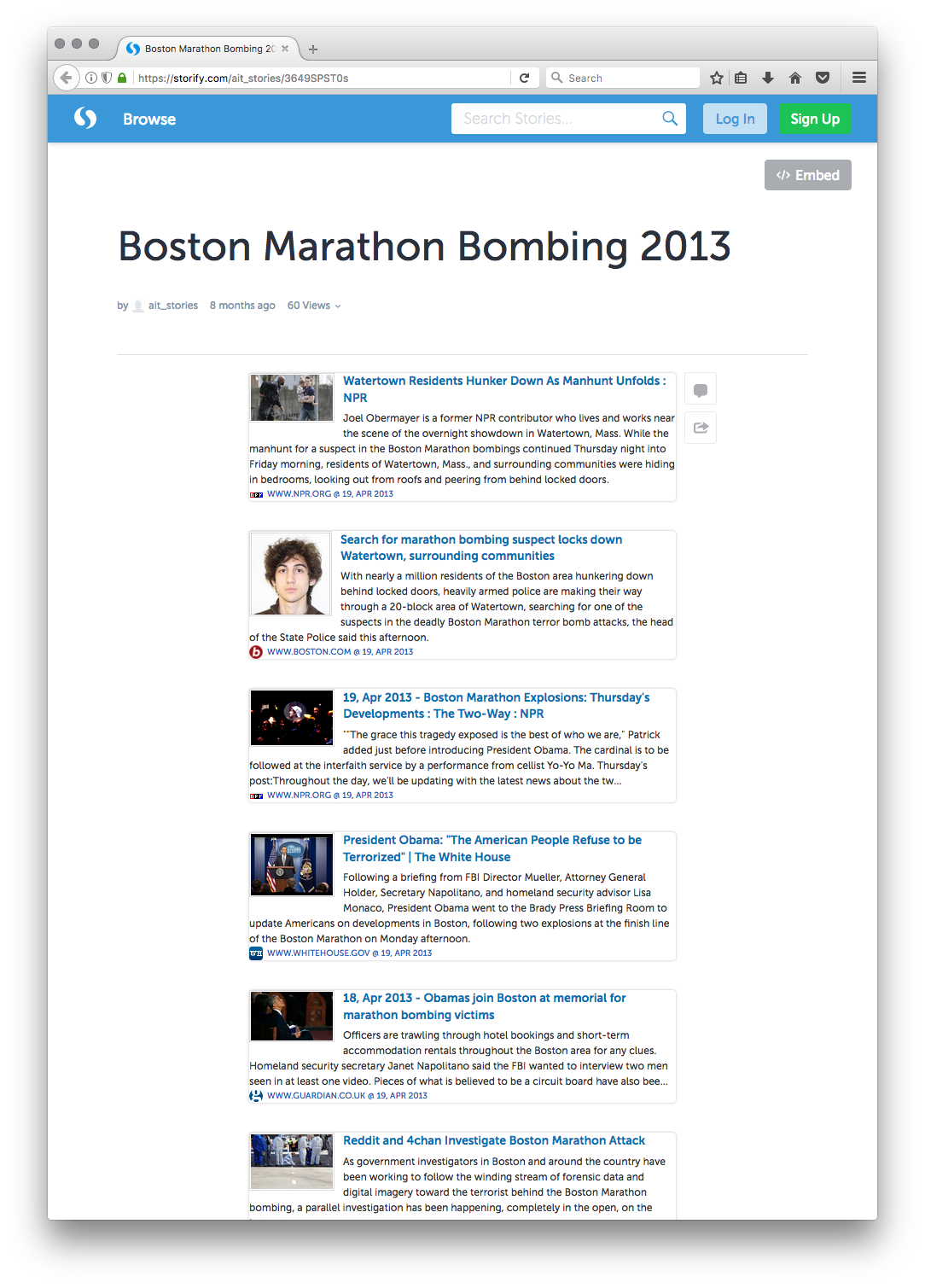}
\caption{An example story summarizing content from the Archive-It web archive collection \emph{Boston Marathon Bombing 2013} using the now-defunct platform Storify.}
\label{fig:storify-example}
\end{figure}

Where a surrogate summarizes a specific document, a user can combine multiple surrogates to create a \textbf{story} summarizing a topic from multiple sources. AlNoamany et al. \cite{alnoamany_generating_2017} was the first to combine this \textbf{storytelling} technique with web archives. Figure \ref{fig:storify-example} shows an example of one of her stories created by combining mementos from the Archive-It collection \emph{Boston Marathon Bombing 2013} and the now-defunct \cite{jones_storify_2017} storytelling service Storify.  Unfortunately, many tools exist for storytelling with web resources, but they have issues working with mementos.

\section{Problems with Combining Existing Storytelling Platforms with Web Archives}
\label{sec:existing_storytelling}

\begin{table}[htbp]
  \begin{tabular}{p{2in} | p{4in}}
  \textbf{Primary Reason Unsuitable for Our Work}                          & \textbf{Curation Platform}                                                       \\ \hline
  No Longer In Service                                                     & Addict-o-matic, AutoEmbed, Bundlr, Google+, Kurator, Newsle, Pulse, Storify                 \\ \hline
  Focus is Engaging with Customers, Not Social Media Storytelling          & Cision, Falcon.io, FlashIssue, Folloze, Sharpr, Spredfast, Sprinklr, Trap!t      \\ \hline
  Content Curation for Website Construction, Not Social Media Storytelling & Curata, CurationSuite, RockTheDeadline, Roojoom, SimilarWeb, Waywire Enterprises \\ \hline
  Social Media Aggregation and Response, Not Social Media Storytelling     & Hootsuite, Pluggio, PostPlanner, SproutSocial                                    \\ \hline
  Focus is Only on Present Content, Not Social Media Storytelling          & ContentGems, DrumpUp, paper.li, The Tweeted Times, Tagboard, UpContent           \\ \hline
  No Public Sharing of Content                                             & Feedly, Pinboard, Pocket, Shareist                                               \\ \hline
  No General URI Support                                                   & Flickr, Huzzazz, Instagram, Togetter, Vidinterest                                \\ \hline
  Surrogate Size Changes, Making Storytelling Difficult                    & Flipboard, Flockler, Juxtapost, Pinterest, Scoop.it                              \\ \hline
  No Public API                                                            & BagTheWeb, ChannelKit, eLink, Listly, Pearltrees, Sutori, Symbaloo               \\ \hline
  No Reliable Surrogate Production                                         & Facebook, Twitter                                                                \\ \hline
  Confuses Information About Memento                                      & Tumblr, Embed.ly, embed.rocks. iframely, noembed, microlink                                                                       
  \end{tabular}
  \caption{The live web curation platforms considered as part of the 2017 review in Section \ref{sec:existing_storytelling} and the reasons as to why they are unsuitable for our storytelling efforts. Embed.ly, embed.rocks. iframely, noembed, microlink were added in 2019.}
  \label{tab:live_web_curation_platforms}
\end{table}

In 2017, we examined a large number of \textbf{live web curation platforms} \cite{jones_where_2017}. In this section, we demonstrate why these platforms are not suitable for our storytelling efforts. Table \ref{tab:live_web_curation_platforms} lists the tools detailed in this section and why they are unsuitable for our work. The list of tools comes from AlNoamany's work \cite{alnoamany_using_2016}, Curata's survey of its potential competitors \cite{hall_content_2017}, and work by Williams \cite{williams_40_2012}. This is a volatile landscape and new companies enter this space frequently. Where Table \ref{tab:live_web_curation_platforms} provides a summary, the rest of this section provides more detail on our findings when using these tools to visualize mementos.

After this review, we came to the conclusion that many of these services would not reliably produce good surrogates for mementos. None provided the level of control AlNoamany exploited to create good surrogates with Storify \cite{alnoamany_generating_2017}. This led us to develop our own surrogate service to support our research efforts.

\subsection{Engaging with Customers, Not Storytelling}

Some tools exist for customer engagement. They provide the ability to curate content from the web to increase confidence in a brand. With Falcon.io, users can share collections internally so that teams can review these collections to craft a message. It allows an organization to curate their own content and coordinate a single message across multiple social channels. They provide social monitoring and analysis of the impact of their message. They use their curated content to develop plans for addressing trends, dealing with crises (e.g., the 2017 Pepsi commercial fiasco \cite{victor_pepsi_2017}), and ensuring that customers know the company is a key player in the market (e.g., IBM's Big Data \& Analytics Hub \cite{curata_ibms_2014}). Cision\footnote{\url{http://www.cision.com/us/}}, FlashIssue\footnote{\url{http://www.flashissue.com/}}, Folloze\footnote{\url{http://folloze.com/}}, Spredfast\footnote{\url{https://www.spredfast.com/}}, Sharpr\footnote{\url{http://sharpr.com/}}, Sprinklr\footnote{\url{https://www.sprinklr.com/}}, and Trap!t\footnote{\url{http://trap.it/}} are tools with a similar focus \cite{zickelbrach_personal_2017,schlachter_personal_2017}. We requested demos and discussions about these tools with similar companies but only received feedback from Falcon.io and Spredfast who were instrumental in helping us understand this space.

Roojoom\footnote{\url{https://www.roojoom.com/}}, Curata\footnote{\url{https://www.curata.com/}}, and SimilarWeb\footnote{\url{https://www.similarweb.com/}}, and Waywire Enterprises\footnote{\url{http://enterprise.waywire.com/}} focus more on helping influence the development of the corporate web site with curated content. RockTheDeadLine\footnote{\url{http://rockthedeadline.com/curation-services/}} offers to curate content on the organization's behalf. CurationSuite\footnote{\url{https://curationsuite.com/pricing/}} (formerly CurationTraffic) focuses on providing a curated collection as a component of a WordPress blog. These services go one step further and provide site integration components in addition to mere content curation.

Hootsuite\footnote{\url{https://hootsuite.com}}, Pluggio\footnote{\url{https://plugg.io}}, PostPlanner\footnote{\url{https://www.postplanner.com}}, and SproutSocial\footnote{\url{https://sproutsocial.com}} focus on collecting and organizing content and responses from social media. They do not provide collections for public consumption in the same way Storify or a Facebook album would. Hootsuite, in particular, provides a way to gather content from many different social networking accounts at once while synchronizing outgoing posts across all of them.

All of these tools offer analytics packages that permit the organization to see how the produced newsletter or web content is performing. Though these tools do focus on curating content, their focus is customer engagement and marketing. Most of these tools focus on trends and web content in aggregate rather than showcasing individual web resources. Our focus  is to find new ways of visualizing representative mementos from web archive collections. Though some of these 
tools might be capable of doing this, their unused excess functionality make them a poor fit for our purpose.

\subsection{Only Displaying the Present}

Some tools allow the user to supply a topic as a seed for curated content. The tool will then use that topic and its internal curation service to locate content that may be useful to the user. A good example is a local newspaper. A resident of Santa Fe, for example, will likely want to know what content is relevant to their city, and hence would be better served by the curation services of the \emph{Santa Fe New Mexican}\footnote{\url{https://www.santafenewmexican.com/}} than they would by the \emph{Seattle Times}\footnote{\url{https://www.seattletimes.com/}}. The newspaper changes every day, but the content reflects the local area.

\begin{figure}[t]
    \centering
    \includegraphics[width=\textwidth]{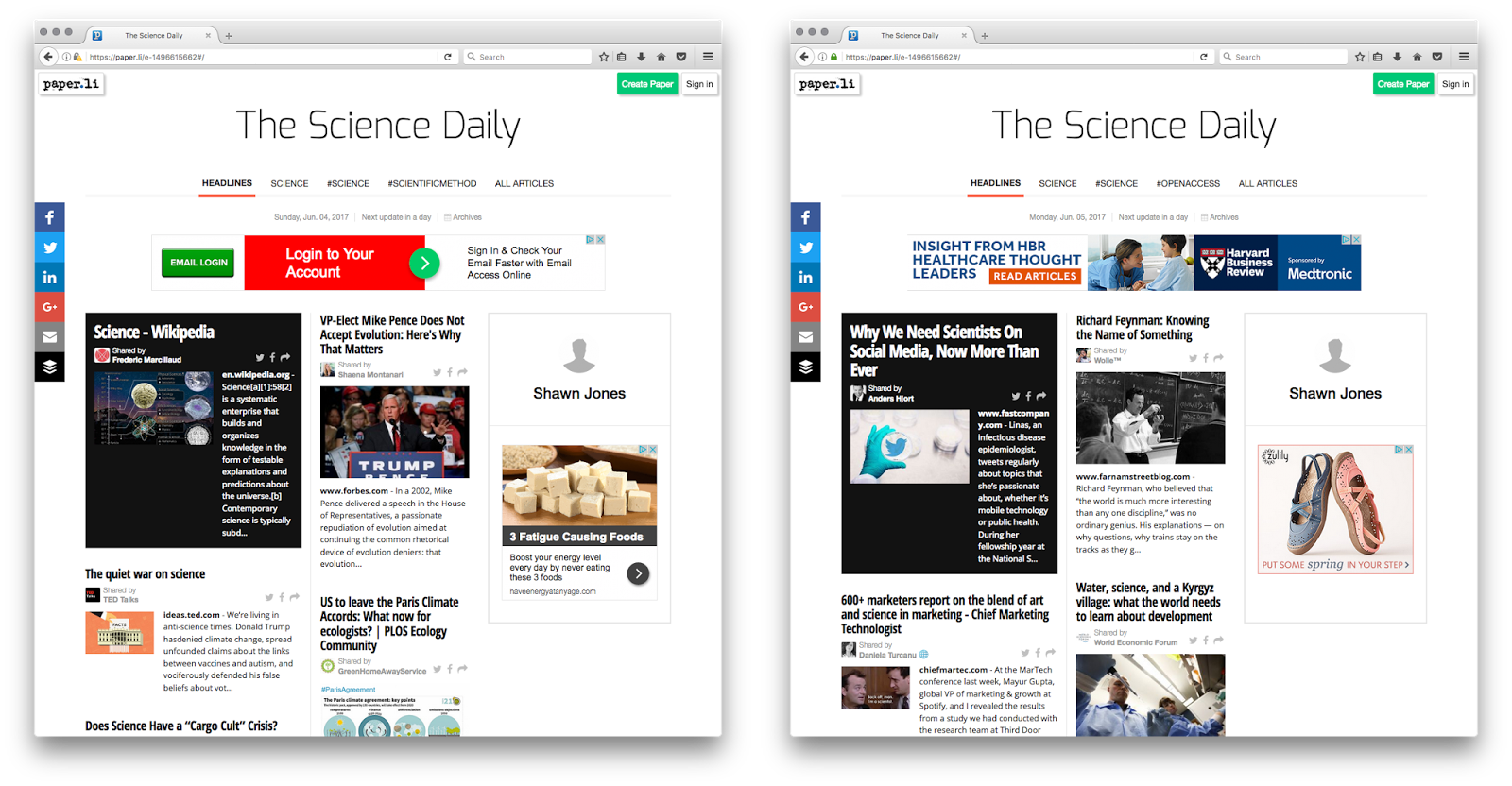}
    \caption[Paper.li presents a different collection each day based on the user's keywords. The user-created \emph{The Science Daily} changes every day. The content for June 4, 2017 (left) is different from the content for June 5, 2017 (right).]{Paper.li presents a different collection each day based on the user's keywords. The user-created \emph{The Science Daily} changes every day. The content for June 4, 2017 (left) is different from the content for June 5, 2017 (right).\\URI: \url{https://paper.li/e-1496615662}}
    \label{fig:paperli_changes}
\end{figure}

\begin{figure}[htbp]
    \centering
    \includegraphics[width=\textwidth]{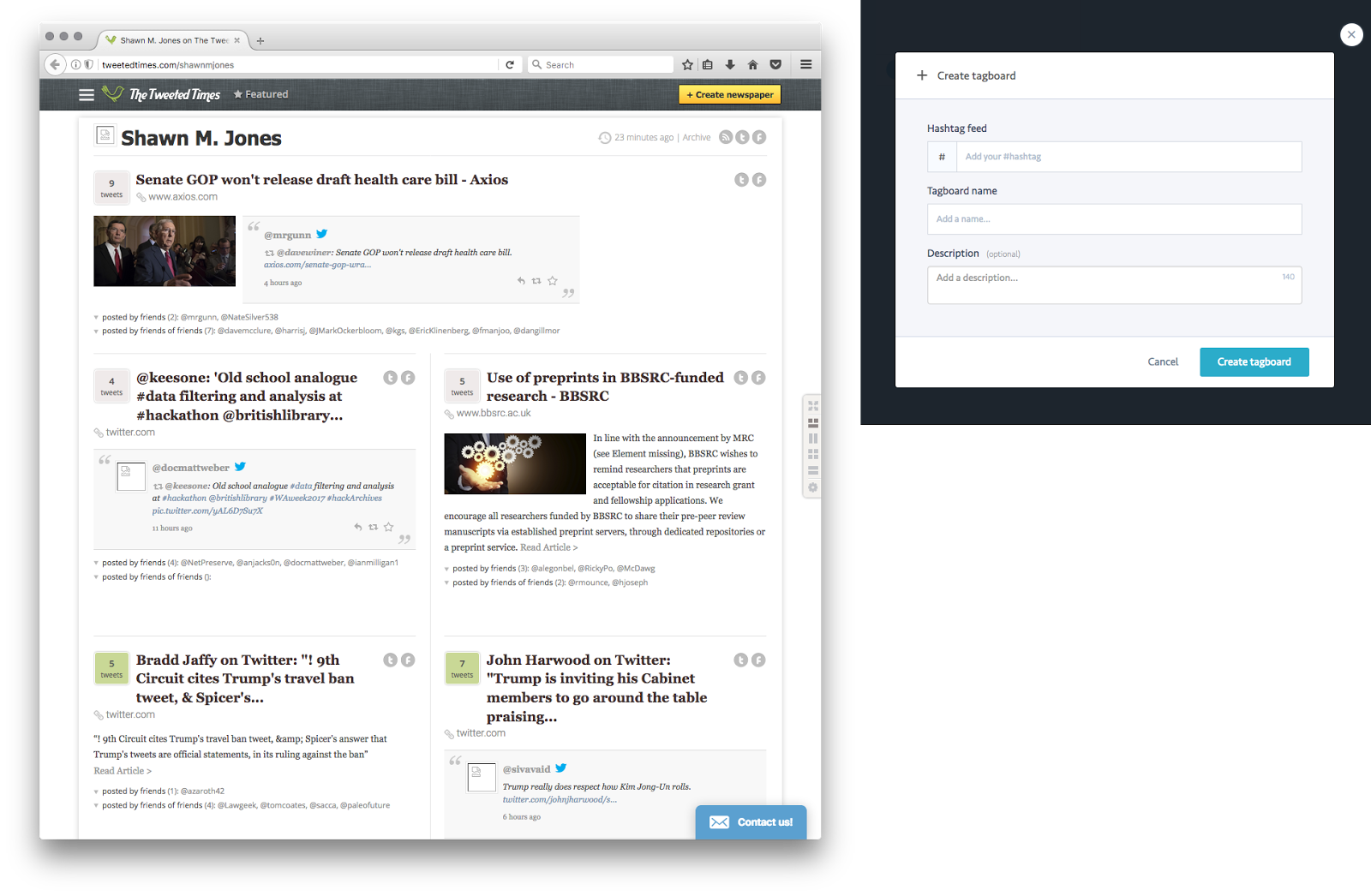}
    \caption{(left) The Tweeted Times shows some of the tweets from the author's Twitter feed. 
    (right) Tagboard requires that a user supply a hashtag as input before creating a collection. }
    \label{fig:tweeted_times-tagboard}
\end{figure}

Geographic location is not the only limiting factor for this category of curation tools. Paper.li\footnote{\url{http://paper.li/}} (Figure \ref{fig:paperli_changes}) and UpContent\footnote{\url{https://upcontent.com/}} allow one to create a personalized newspaper about a given topic that changes each day, providing fresh content to the user. ContentGems\footnote{\url{https://contentgems.com/}} is much the same but supports a complex workflow system that can be edited to supply content from multiple sources. ContentGems also allows one to share their generated paper via email, Twitter, RSS feed, website widgets, RSS, IFTTT\footnote{\url{https://ifttt.com/}}, Zapier\footnote{\url{https://zapier.com/}}, and a whole host of other services. DrumUp\footnote{\url{https://drumup.io/}} uses a variety of sources from the general web and social media to generate topic-specific collections. They also allow the user to schedule social media posts to Facebook, Twitter, and LinkedIn. Paper.li appears to be focused on a single user, ContentGems and DrumUp easily stretch into customer engagement, and UpContent offers different capabilities depending on to which tier the user has subscribed.

The Tweeted Times\footnote{\url{http://tweetedtimes.com/}} and Tagboard\footnote{\url{https://tagboard.com/}} (Figure \ref{fig:tweeted_times-tagboard}) both focus on content from social media. The Tweeted Times attempts to summarize a user's Twitter feed and later publishes that summary at a URI for the end user to consume. Tagboard uses hashtags from Facebook or Twitter as seeds to their content curation system.

The tools in this section focus on content from the present. They do not allow a user to supply a list of URIs and view them as surrogates, hence are not suitable for our storytelling efforts.

\subsection{Poor Sharing or No Social Cards}

There is a spectrum of sharing. Storify allowed one to share their collection publicly for all to see. Other tools expect only subscribed accounts to view their collections. In these cases, subscribed accounts may be acquired for free or at cost. Feedly\footnote{\url{https://feedly.com/i/welcome}} supports the sharing of collections only for other users in one's team, a grouping of users that can view each other's content. Pinboard\footnote{\url{http://pinboard.in/}} and Pocket\footnote{\url{https://getpocket.com/}} are slightly less restrictive, permitting other portal users to view their content. Also, both Pinboard and Pocket promise paying customers the ability to archive their saved web resources for later viewing. Shareist\footnote{\url{http://www.shareist.com/}} only shares content via email and on social media, but does not produce a web-based visualization of a collection. We are interested in tools that allow us to not only share collections of mementos on the Web but also share them with as few barriers to viewing as possible.

\begin{figure}[t]
    \centering
    \includegraphics[width=0.6\textwidth]{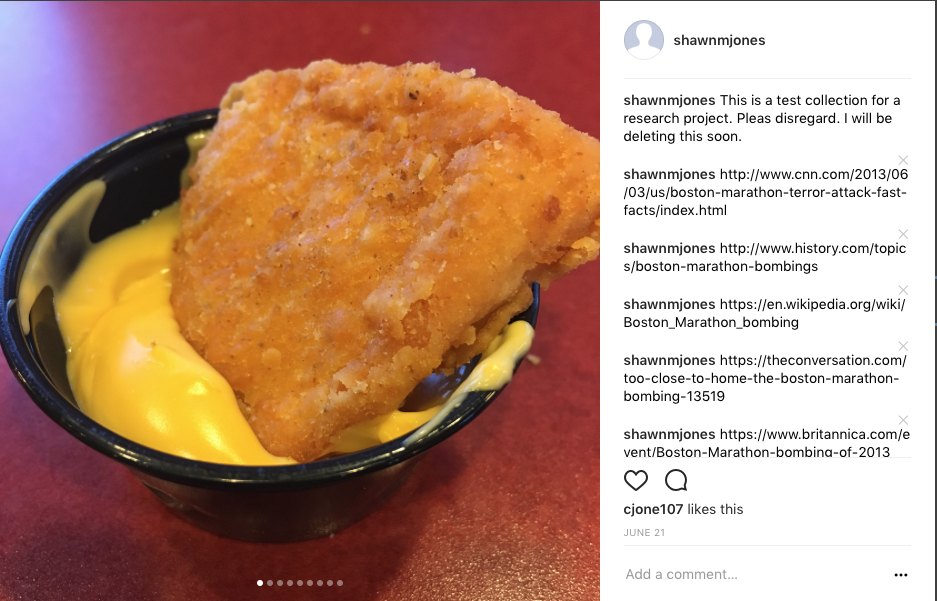}
    \caption{An example Instagram post. Note that the links in the comment are not converted to surrogates, or even clickable text.}
    \label{fig:instagram_example}
\end{figure}

\begin{figure}[htbp]
    \begin{subfigure}[t]{\textwidth}
        \centering
        \includegraphics[width=0.7\textwidth]{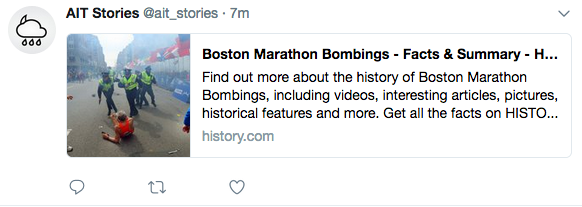}
        \caption{A Tweet displaying a social card.\\URI: \url{https://twitter.com/ait_stories/status/882730783949168640}}
        \label{fig:test_tweet_with_socialcard}
    \end{subfigure}
    \par\bigskip
    \begin{subfigure}[t]{\textwidth}
        \centering
        \includegraphics[width=0.7\textwidth]{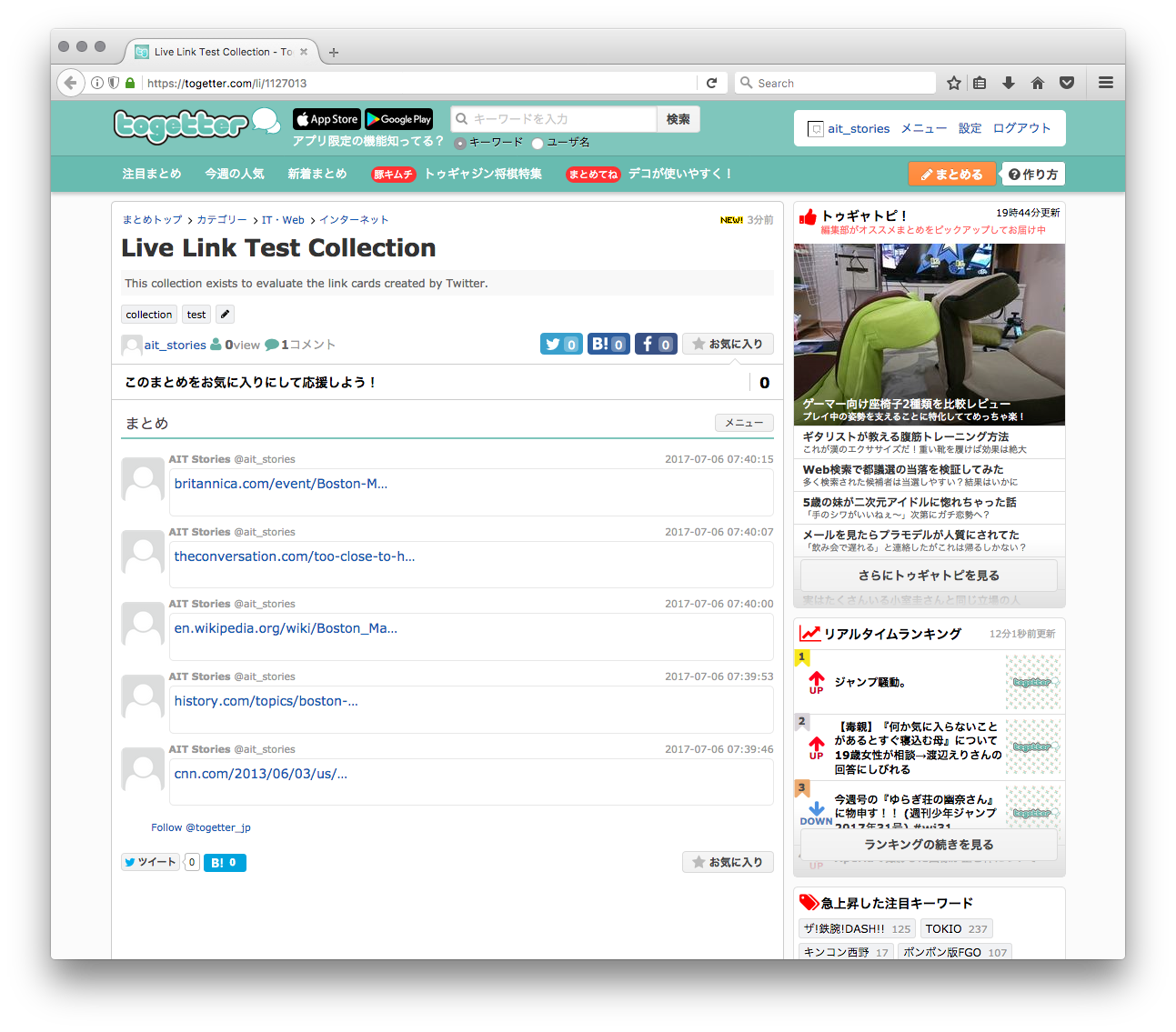}
        \caption{A Togetter collection containing the Tweet from Figure \ref{fig:test_tweet_with_socialcard}.\\URI: \url{https://togetter.com/li/1127013}}
    \end{subfigure}
    \caption{Togetter is not suitable for our purposes because it does not display social cards, even though they are displayed in the Tweets making up a collection.}
    \label{fig:togetter_issues}
\end{figure}

Huzza\footnote{\url{https://huzzaz.com}} and Vidinterest\footnote{\url{https://vidinterest.tv}} only support URIs to web resources that contain video. Both support YouTube and Vimeo URIs, but only Vidinterest supports Dailymotion. Neither support general URIs, let alone URI-Ms. Instagram\footnote{\url{https://www.instagram.com}} and Flickr\footnote{\url{https://www.flickr.com}} work specifically with images, and they do not create social cards for URIs (Figure \ref{fig:instagram_example}). As shown in Figure \ref{fig:togetter_issues}, even though Twitter may render a social card in the Tweets, the card is not present when the tweets are visualized in a collection using Togetter\footnote{\url{https://togetter.com}}.

The lack of surrogates and the limitations on sharing and media types make these tools unsuitable for our storytelling efforts.

\subsection{Confusing Story Flow}

Some tools change the size of the card for effect, or to allow extra data in one card rather than another. These size changes interrupt the visual flow of the small multiples paradigm that makes storytelling successful. While good for presenting in newspapers or other tools that collect articles, such size changes make it difficult to follow the flow of events in a story. They create additional cognitive load on the user, forcing her to regularly ask ``does this different sized card come before or after the other cards in my view?'' and ``how does this card fit into the story timeline?''

\begin{figure}[htbp]
    \centering
    \includegraphics[width=0.7\textwidth]{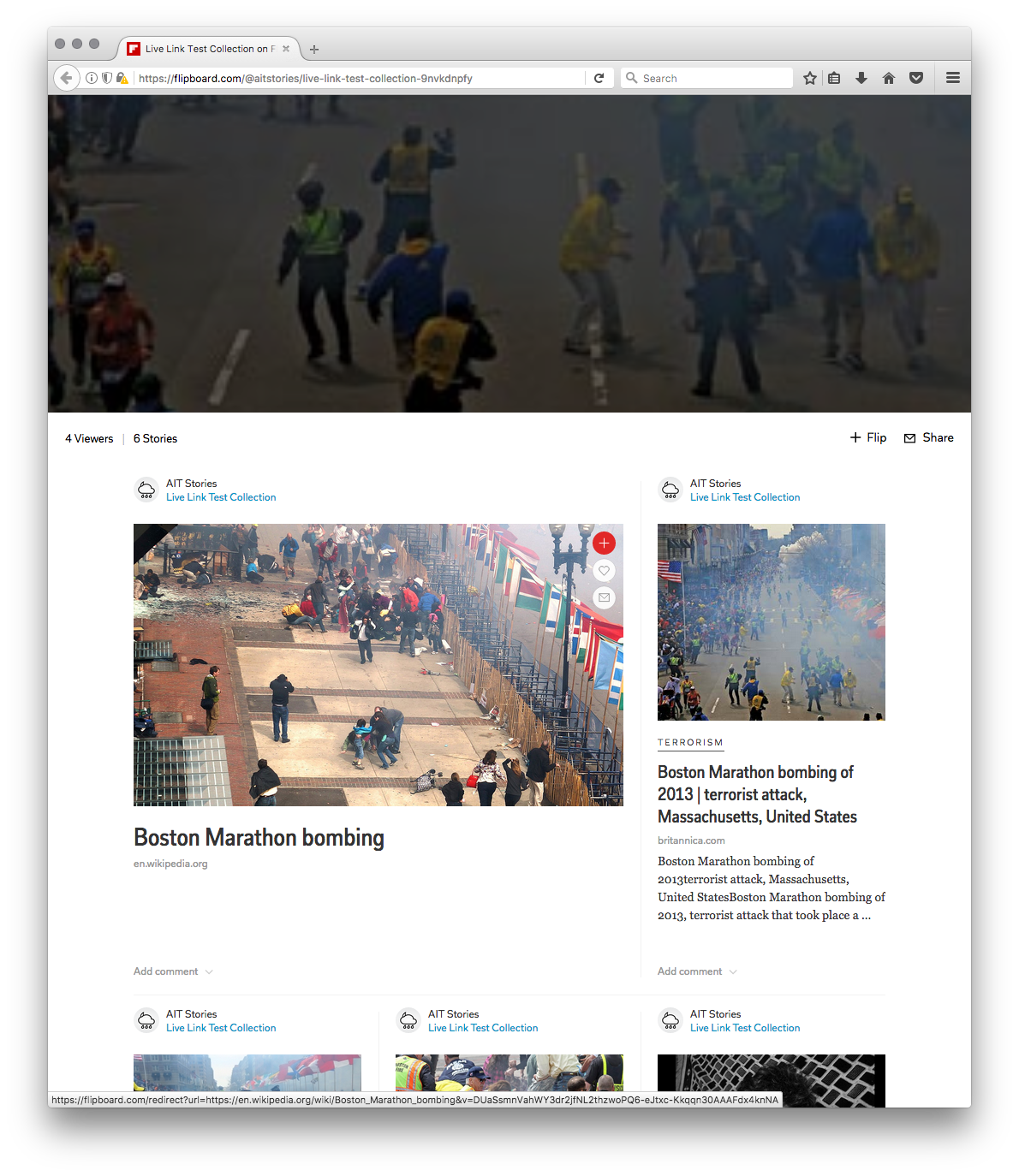}
    \caption{Flipboard orders the social cards from left to right then up and down, but changes the size of some of the cards.}
    \label{fig:flipboard_example}
\end{figure}

\begin{figure}[htbp]
    \centering
    \includegraphics[width=0.7\textwidth]{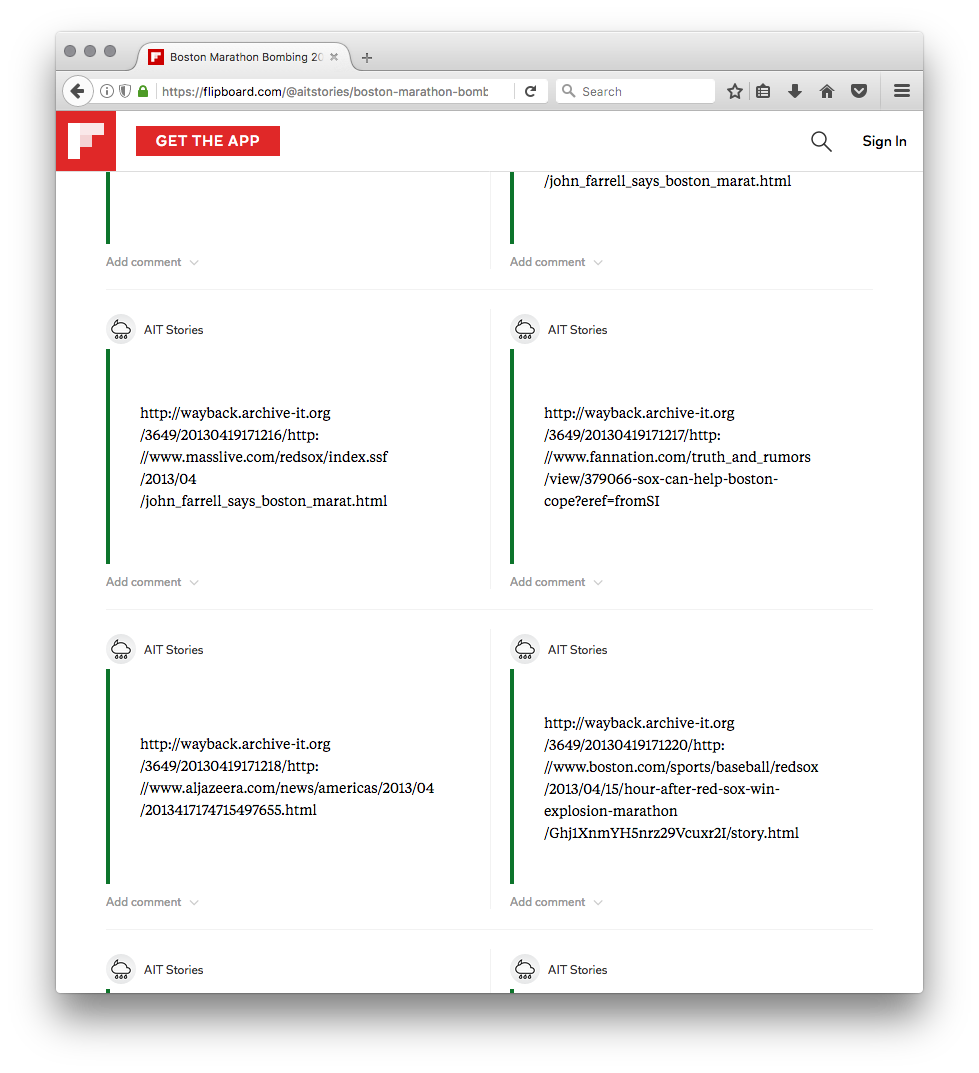}
    \caption{Unlike live web resources, Flipboard does not render mementos as social cards.}
    \label{fig:flipboard_mementos}
\end{figure}

Flipboard\footnote{\url{https://flipboard.com}} often makes the first social card the largest, dominating the collection as seen in Figure \ref{fig:flipboard_example}. Sometimes it will choose another card in the collection and increase its size as well. Flipboard also has other issues. In Figure \ref{fig:flipboard_example}, we see social cards rendered for live links, but in Figure \ref{fig:flipboard_mementos} we see that Flipboard does not do so well with mementos.

\begin{figure}[htbp]
    \centering
    \includegraphics[width=0.7\textwidth]{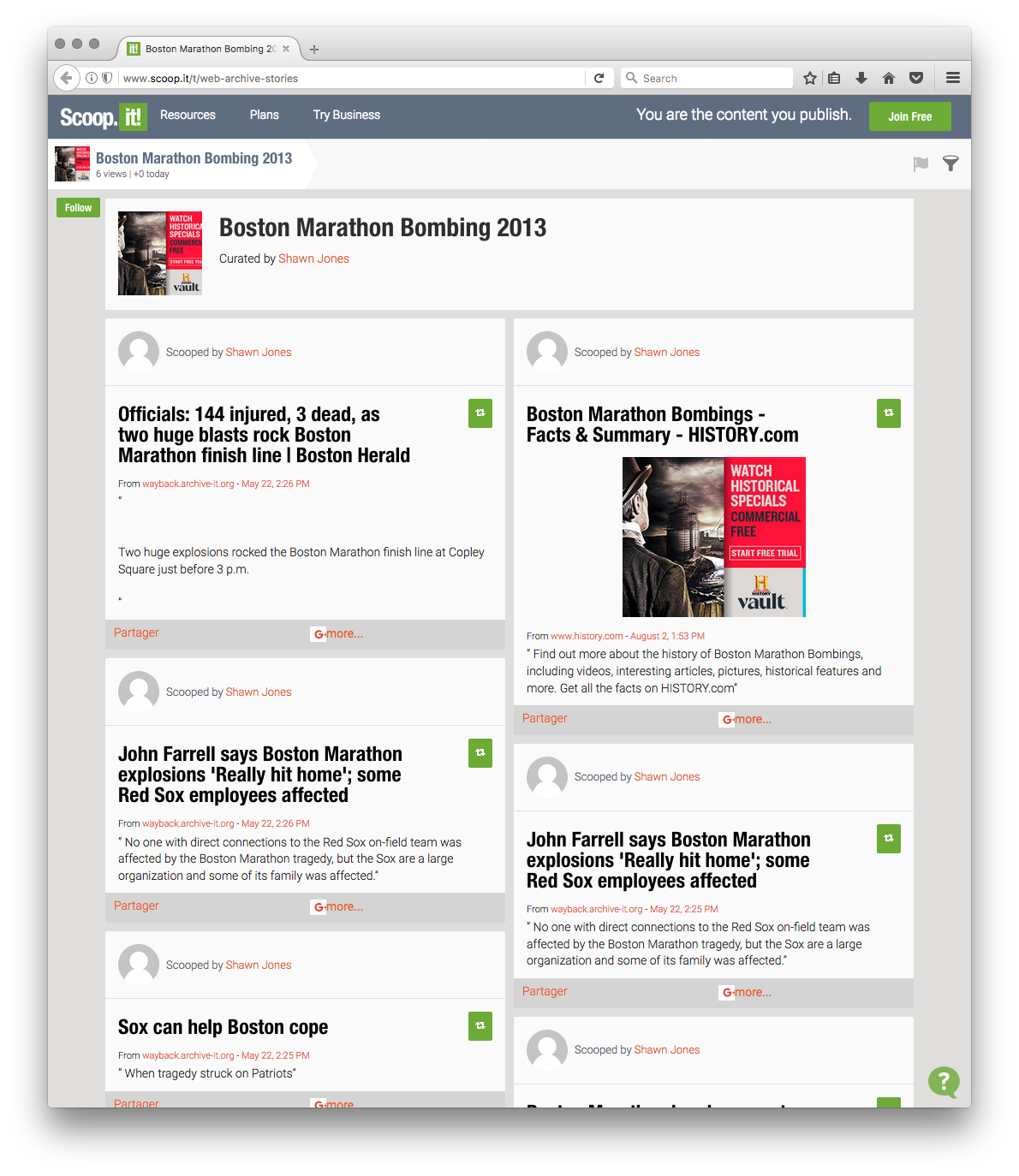}
    \caption[In this Scoop.it collection, Scoop.it changes the size of some social cards based on the amount of data present in the card.]{In this Scoop.it collection, Scoop.it changes the size of some social cards based on the amount of data present in the card.\\URI: \url{https://www.scoop.it/topic/web-archive-stories}}
    \label{fig:scoopit_example}
\end{figure}

Scoop.it\footnote{\url{https://www.scoop.it}} (Figure \ref{fig:scoopit_example}) changes the size of some social cards due to the presence of large images or extra text in the snippet. These changes distort the visual flow of the collection. There are also restrictions, even for paying users, on the amount of content that users can store, with users being limited to only 15 collections, even though they subscribed to the top \$33 per month subscription.

\begin{figure}[htbp]
    \centering
    \includegraphics[width=0.7\textwidth]{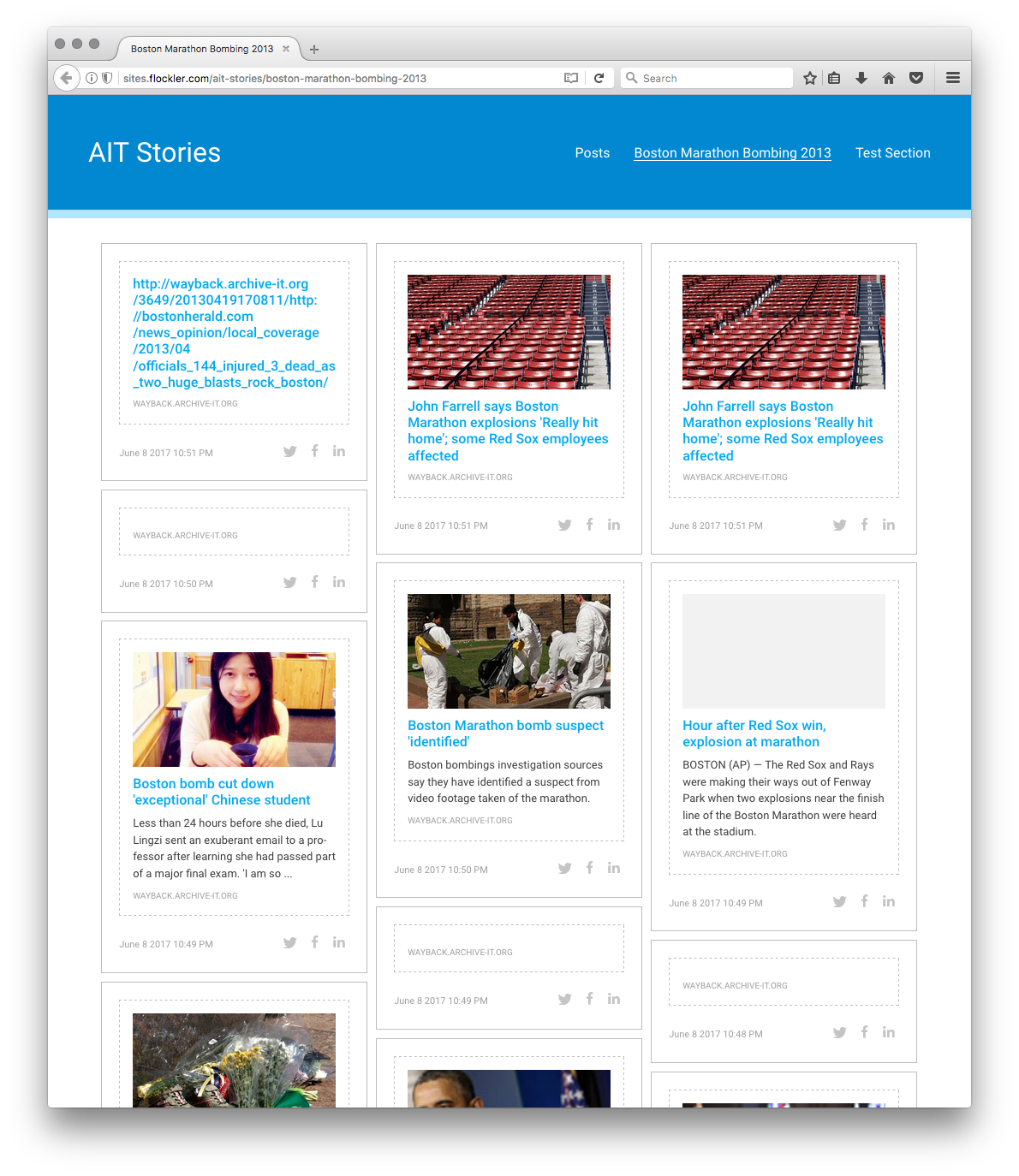}
    \caption{Flockler alters the sizes of some cards based on the information present. It also fails to generate cards for some mementos.}
    \label{fig:flockler_example}
\end{figure}

Flockler\footnote{\url{https://flockler.com}} alters the size of its cards based on the information present. Cards with images, titles, and snippets are larger than those with just text. As shown in Figure \ref{fig:flockler_example}, sometimes Flockler cannot extract any information and generates blank cards or cards whose title is the URI.

\begin{figure}[htbp]
    \centering
    \includegraphics[width=0.7\textwidth]{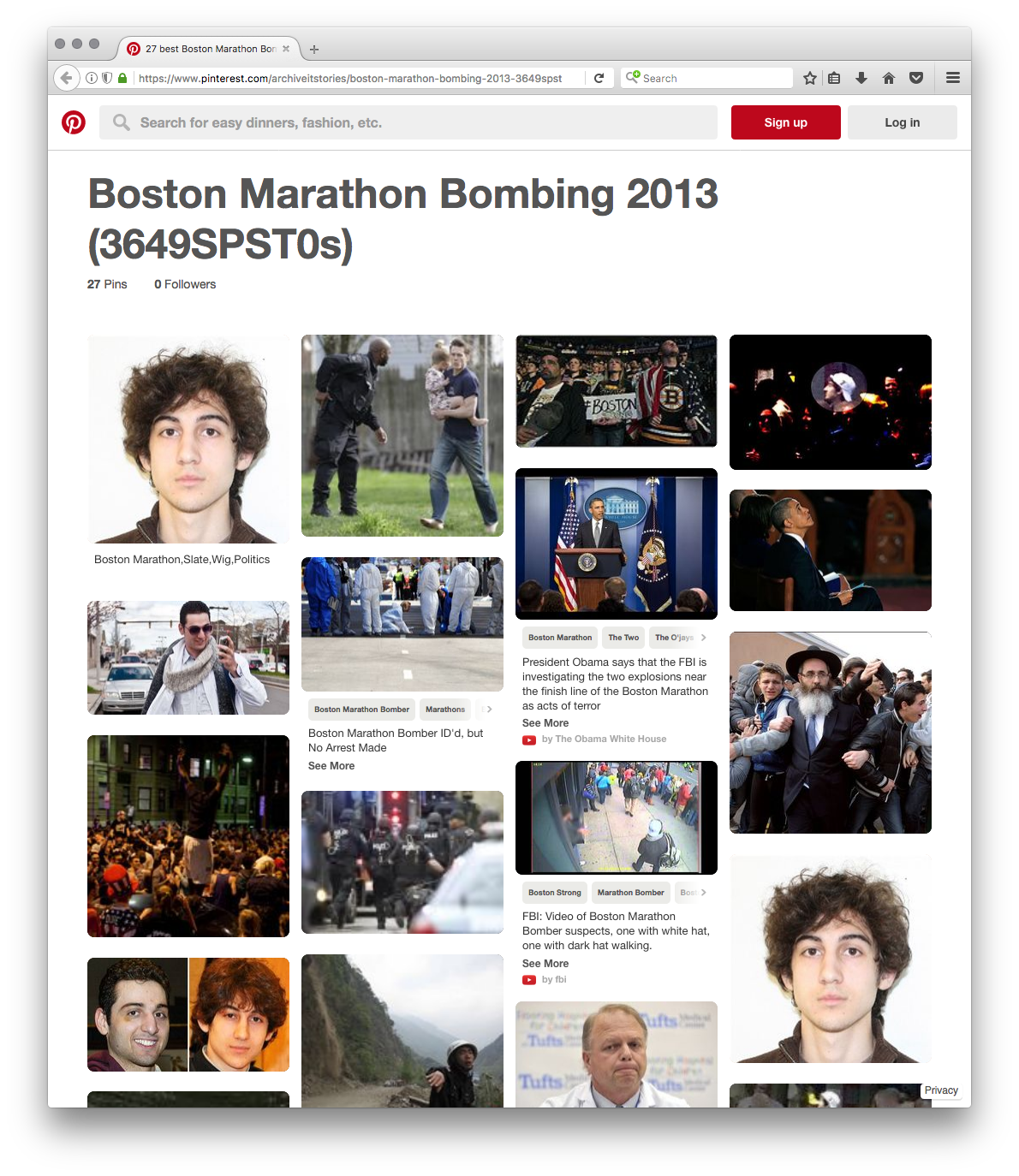}
    \caption[Pinterest supports collections, but does not typically generate social cards, favoring images as seen in this example.]{Pinterest supports collections, but does not typically generate social cards, favoring images as seen in this example.\\URI: \url{https://www.pinterest.com/archiveitstories/boston-marathon-bombing-2013-3649spst0s/}}
    \label{fig:pinterest_screenshot}
\end{figure}

Pinterest has a distinct focus on images but does create social cards  (i.e., ``pins'' in the Pinterest nomenclature) for web resources. The system requires a user to select an image for each pin. Unfortunately, the images are all different sizes (Figure \ref{fig:pinterest_screenshot}), making it difficult to follow the sequence of events in the story. In addition to the size issue, if Pinterest cannot find an image in a page or if the image is too small, it will not create a surrogate. This lack of consistency means that Pinterest is not reliable enough for our storytelling efforts.

\begin{figure}[htbp]
    \centering
    \includegraphics[width=0.7\textwidth]{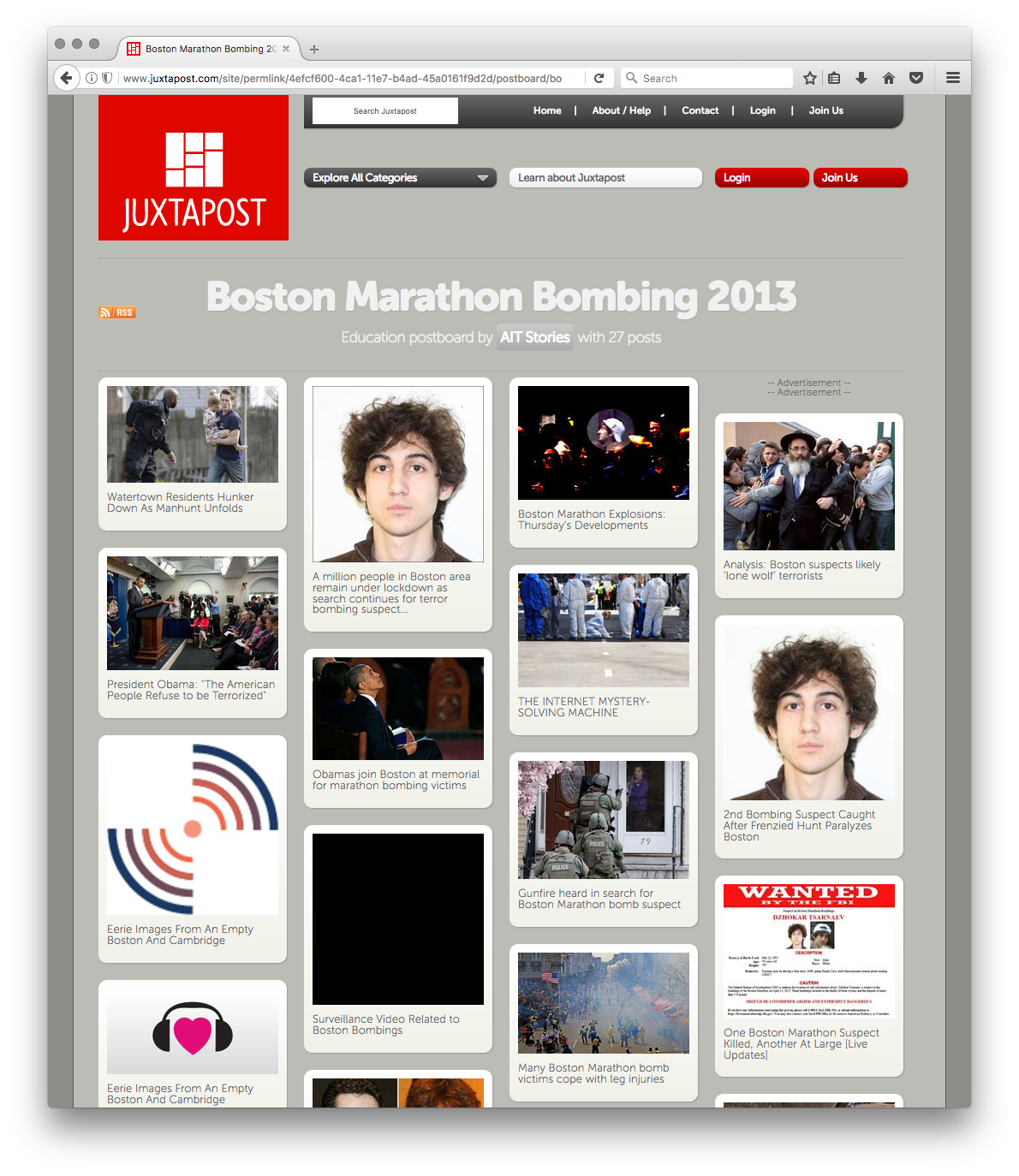}
    \caption[As seen in this collection, Juxtapost changes the size of social cards and even moves them out of the way for advertisements (top right text says ``--Advertisement--''). Which direction does the story flow?]{As seen in this collection, Juxtapost changes the size of social cards and even moves them out of the way for advertisements (top right text says ``--Advertisement--''). Which direction does the story flow?\\URI: \url{http://www.juxtapost.com/site/permlink/4efcf600-4ca1-11e7-b4ad-45a0161f9d2d/postboard/boston_marathon_bombing_2013/}}
    \label{fig:juxtapost_example}
\end{figure}

Juxtapost\footnote{\url{http://www.juxtapost.com}} (Figure \ref{fig:juxtapost_example}) is the other tool which changes the size of the social cards. Like with Pinterest, Juxtapost requires that the end user select an image and insert a description for every card. 

\subsection{No API}

Our research for summarizing collections requires that our system generate the story automatically. We require a web API. Would it be acceptable for a human to submit these links from our solution to one of these tools? What if the collection changes frequently and our solution must be executed again to account for these changes?

Pearltrees has no API\footnote{\url{http://www.pearltrees.com/s/faq/en\#Q.4.1.4}}.  We could not locate documentation of APIs for Symbaloo\footnote{\url{https://www.symbaloo.com/}}, eLink\footnote{\url{https://elink.io/}}, ChannelKit\footnote{\url{https://www.channelkit.com/landing/home}}, or BagTheWeb\footnote{\url{http://www.bagtheweb.com/}}. Both Sutori \cite{ketchell_personal_2018} and Listly have APIs, but neither are public\footnote{\url{https://community.list.ly/t/where-to-find-the-api-documentation/294}}.

BagTheWeb requires additional information supplied by the user in order to create a social card. BagTheWeb does not generate any social card data based solely on the URI. If there were an API, our solution might be able to address some of these shortcomings. Symbaloo is much the same. It chooses an image but often favors the favicon over an image selected from the article.

\begin{figure}[t]
    \centering
    \includegraphics[width=0.6\textwidth]{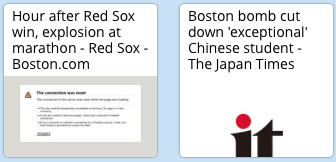}
    \caption[A screenshot of two social cards created from Archive-It mementos by Pearltrees in a collection about the Boston Marathon. The one on the left displays a Firefox error instead of a selected image for the memento.]{A screenshot of two social cards created from Archive-It mementos by Pearltrees in a collection about the Boston Marathon. The one on the left displays a Firefox error instead of a selected image for the memento.\\URI: \url{http://www.juxtapost.com/site/permlink/4efcf600-4ca1-11e7-b4ad-45a0161f9d2d/postboard/boston_marathon_bombing_2013/}}
    \label{fig:pearltrees_example}
\end{figure}

Pearltrees has problems that may be addressed by an API that allows the user to specify information. Figure \ref{fig:pearltrees_example} displays a Firefox error instead of a selected image in the social card. This image is especially surprising because the system was able to extract the title from the destination URI. Pearltrees also tends to convert URI-Ms to URI-Rs, linking to the live page instead of the memento.

\begin{figure}[htbp]
    \centering
    \includegraphics[width=0.6\textwidth]{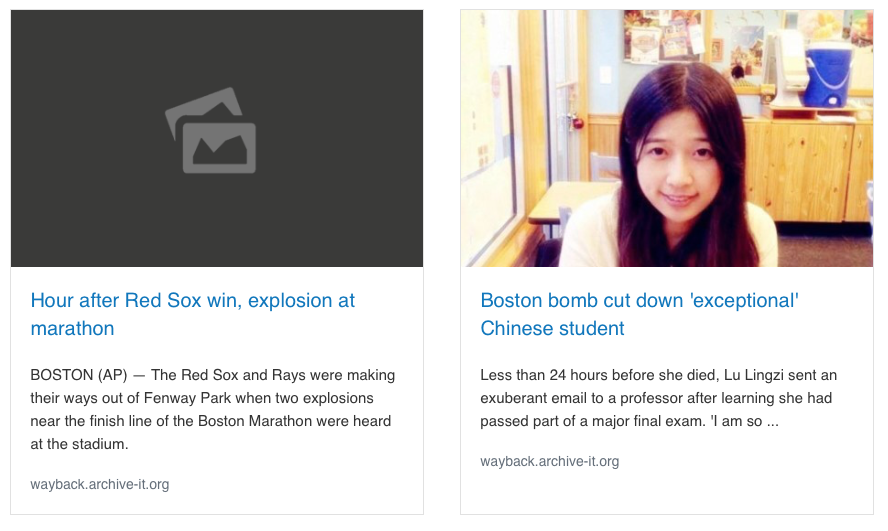}
    \caption[A screenshot of two social cards generated from Archive-It mementos from an eLink collection about the Boston Marathon Bombing. The one of the left shows a missing selected image while the one on the right displays fine.]{A screenshot of two social cards generated from Archive-It mementos from an eLink collection about the Boston Marathon Bombing. The one of the left shows a missing selected image while the one on the right displays fine.\\URI: \url{https://elink.io/p/boston-marathon-bombing-2013}}
    \label{fig:elink_example}
\end{figure}

The social cards generated by eLink have a selected image, a title, and a text snippet. Sometimes, however, eLink does not seem to find an image on the page. Scoop.it also has similar problems for some URIs. An API call that allows one to select an image for the card would help improve these tools' shortcomings.

\begin{figure}[htbp]
    \centering
    \includegraphics[width=0.6\textwidth]{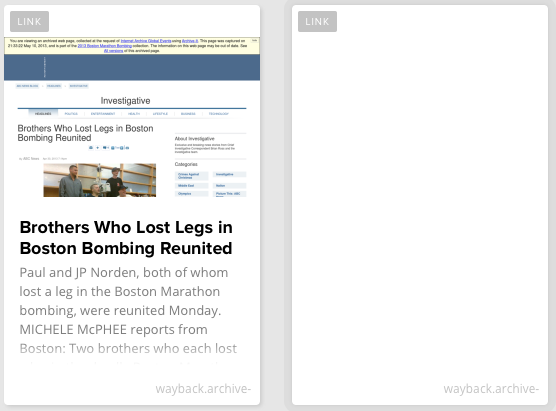}
    \caption[A screenshot of the social cards generated from Archive-It mementos in a ChannelKit collection about the Boston Marathon Bombing. The one on the right shows no useful information.]{A screenshot of the social cards generated from Archive-It mementos in a ChannelKit collection about the Boston Marathon Bombing. The one on the right shows no useful information.\\URI: \url{https://www.channelkit.com/ait_stories/boston-marathon-bombing-2013}}
    \label{fig:channelkit_example}
\end{figure}

\begin{figure}[htbp]
    \centering
    \includegraphics[width=0.6\textwidth]{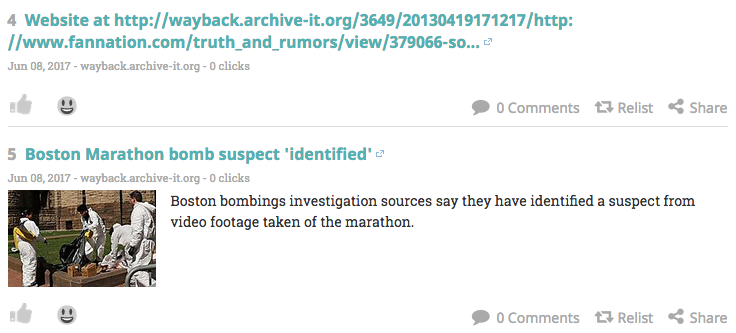}
    \caption[A screenshot of social cards generated from Archive-It mementos in a Listly collection about the Boston Marathon Bombing. The one on the top has no information but the URI. The one on the bottom contains a title, selected image, and snippet.]{A screenshot of social cards generated from Archive-It mementos in a Listly collection about the Boston Marathon Bombing. The one on the top has no information but the URI. The one on the bottom contains a title, selected image, and snippet.\\URI: \url{https://list.ly/list/1XZ9-boston-marathon-bombing-20135}}
    \label{fig:listly_example}
\end{figure}

ChannelKit's social cards are complete with a title, text snippet, and a selected image or web page thumbnail. Sometimes, as seen in Figure \ref{fig:channelkit_example}, the resulting card contains no information, and a human must intervene. As shown in Figure \ref{fig:listly_example}, Listly also has issues with some of the links submitted to it. It usually generates a title, text snippet, and selected image, but in some cases, just lists the URI. Flockler also has similar problems. An API call that allows one to supply the missing information would help address us these issues.

\subsection{Poor Performance From Remaining Platforms}

\begin{figure}[t]
    \centering
    \includegraphics[width=0.7\textwidth]{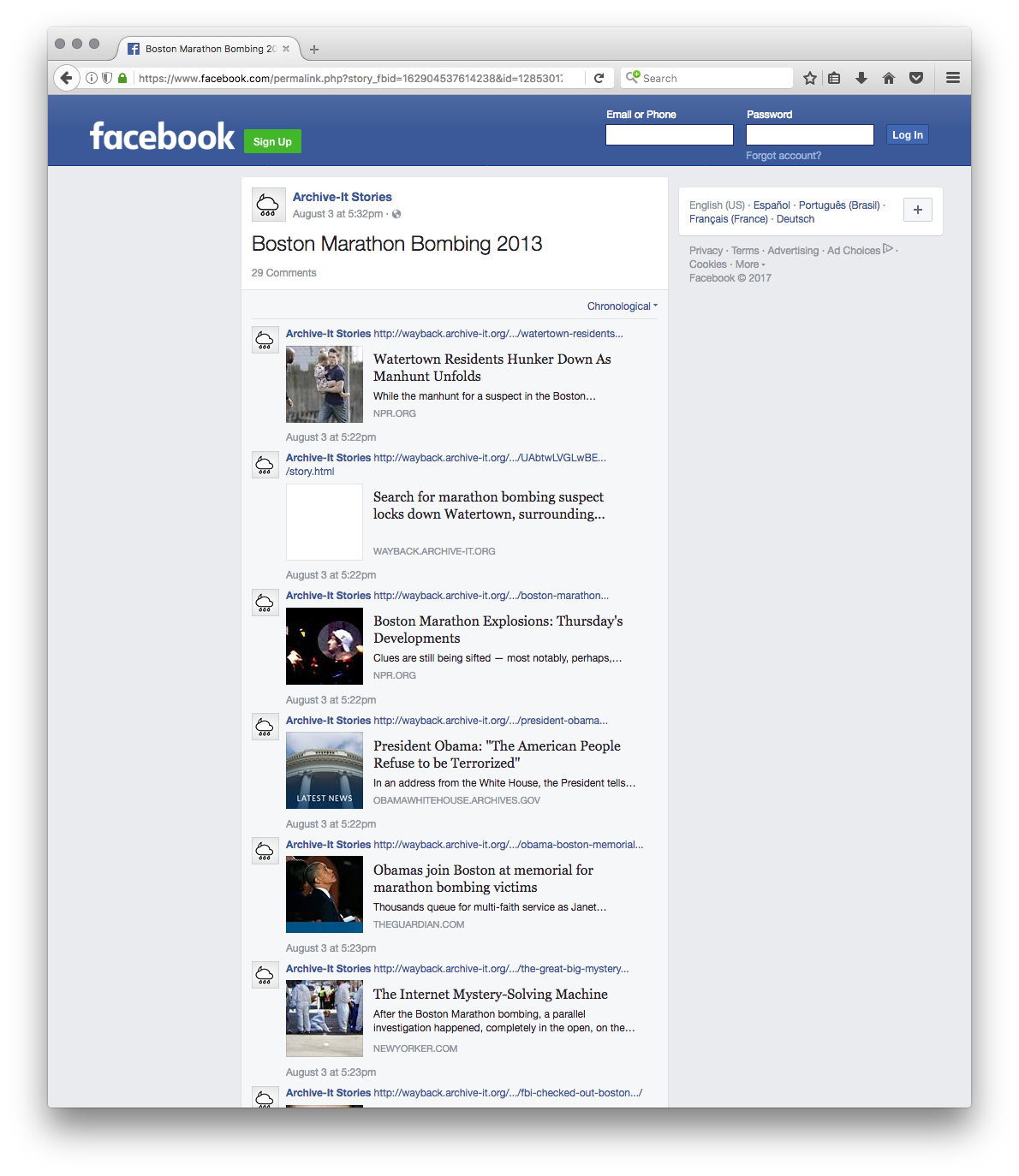}
    \caption[Selected mementos from Archive-It collection 3649 about the Boston Marathon Bombing as visualized in social cards in Facebook comments where the collection is stored as a Facebook post.]{Selected mementos from Archive-It collection 3649 about the Boston Marathon Bombing as visualized in social cards in Facebook comments where the collection is stored as a Facebook post.\\URI: \url{https://www.facebook.com/permalink.php?story_fbid=162904537614238&id=128530177718341}.}
    \label{fig:facebook_collection}
\end{figure}

With 2.2 billion users \cite{disparte_facebook_2018}, Facebook is the most popular social media tool. Facebook supports social cards in posts and comments. Facebook also supports creating albums of photos, but not of posts. Posts contain comments, however. In order to generate a series of social cards in a collection, we gave the post the title of the collection and supplied each URI-M in the story to a separate comment, as seen in Figure \ref{fig:facebook_collection}. In this way, we can generate a story.

\begin{figure}[htbp]
    \centering
    \includegraphics[width=0.8\textwidth]{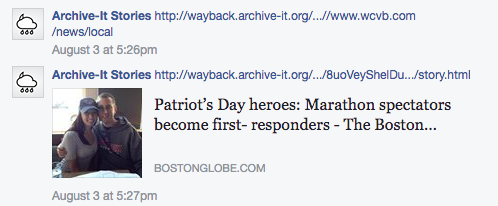}
    \caption{A screenshot of two Facebook comments. The URI-M below generated a social card, but the URI-M above did not.}
    \label{fig:facebook_socialcard}
\end{figure}

As seen in Figure \ref{fig:facebook_socialcard}, Facebook does occasionally fail to generate social cards for links. The Facebook API\footnote{\url{https://developers.facebook.com/}} could be used to update such comments with a photo and a snippet, if necessary. Providing additional images is not possible, as Facebook posts and comments will not generate a social card if the post/comment already has an image. To ensure that we maximize the social card capability of Facebook, we have two options. We can generate all text snippets and select images ourselves for each memento. Alternatively, we could test to see which URI-Ms fail to produce a social card. In the case of either, we would be generating surrogates ourselves and not really using Facebook for anything more than a publishing medium.

\begin{figure}[htbp]

  \centering

  \begin{subfigure}[t]{0.45\textwidth}
    \centering
    \includegraphics[width=\textwidth]{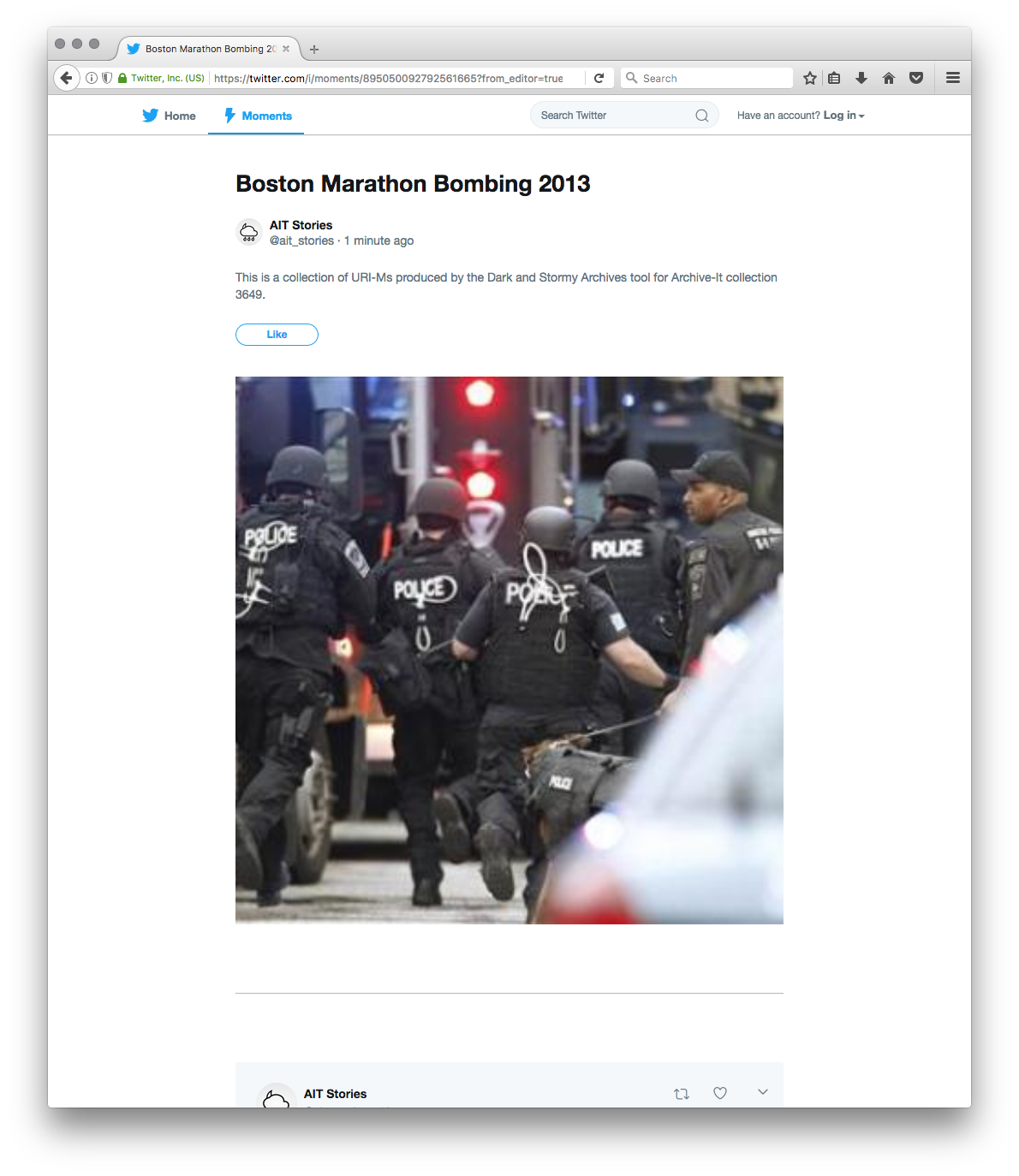}
    \caption{We must manually upload our own photo for this Twitter Moment}
    \label{fig:twitter_moments_example}
  \end{subfigure}%
  ~
  \begin{subfigure}[t]{0.45\textwidth}
      \centering
      \includegraphics[width=\textwidth]{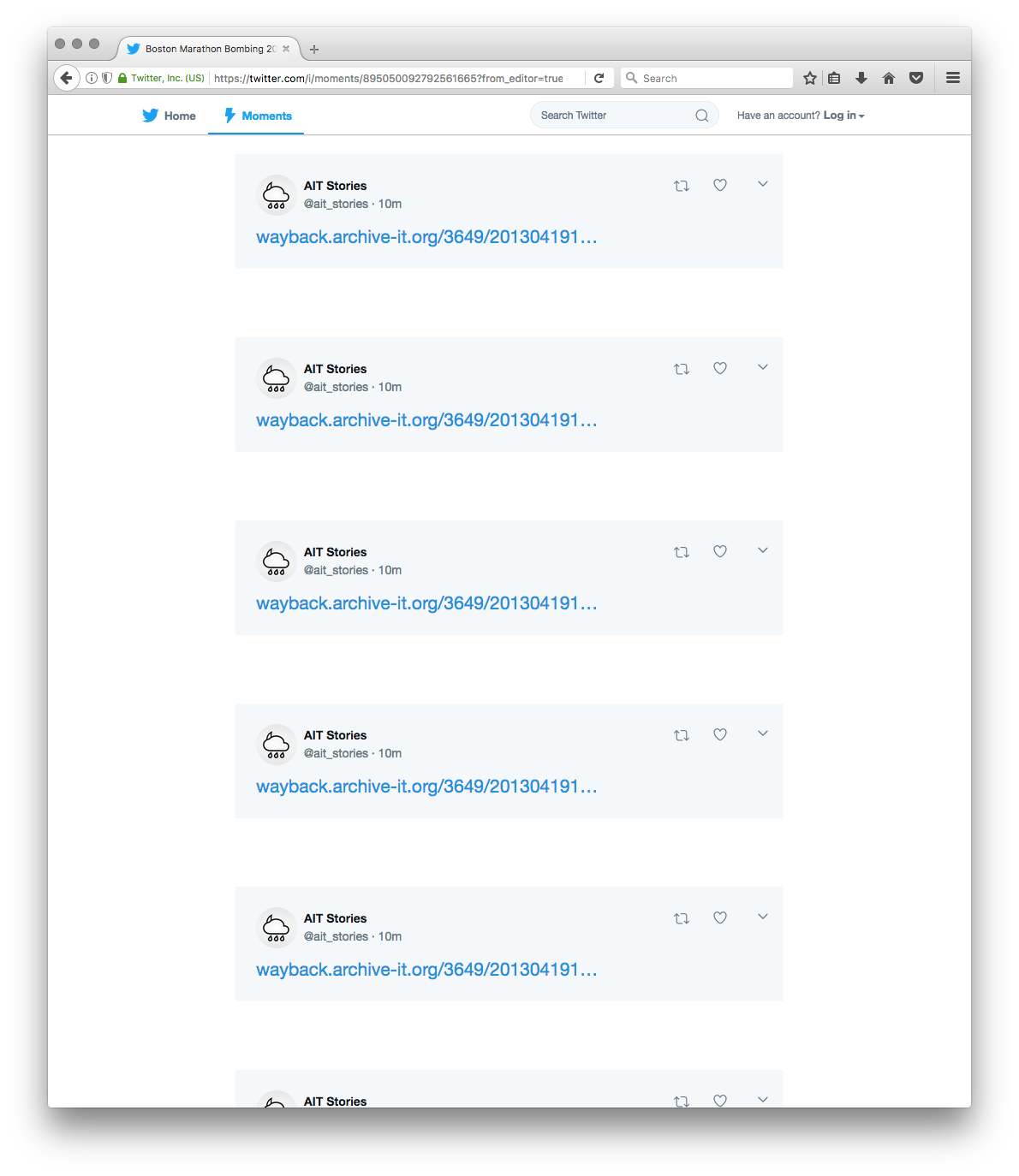}
      \caption{Tweets in the Twitter Moment do not render cards for our Archive-It URI-Ms}
      \label{fig:twitter_moments_no_cards}
  \end{subfigure}

  \caption{This Twitter Moment contains tweets that contain the URI-Ms from our Dark and Stormy Archives summary.}
  
\end{figure}

While all tools required that the user title the story in some way, as shown in Figure \ref{fig:twitter_moments_example}, Twitter Moments requires that the user upload an image separately in order to create a story. This image serves as the striking image for the story. The user is also compelled to supply a description. Sadly, much like Flipboard, Twitter Moments does not appear to generate social cards for URI-Ms. Shown in Figure \ref{fig:twitter_moments_no_cards}, Twitter Moments display URI-Ms with no additional visualization.

For those tweets that fail to produce social cards, we could use the Twitter API to add images and additional text (up to 280 characters of course) to supplement these tweets. At that point, we are building our own social cards out of tweets and just using Twitter as a publishing medium, just like we would be if we had to work around Facebook's reluctance to produce social cards in some cases. We will highlight how Raintale accomplishes this in Section \ref{sec:raintale}.

\subsection{Confusing Archive Data With Content}
\label{sec:card_services}

\begin{figure}
\centering
\includegraphics[width=0.6\textwidth]{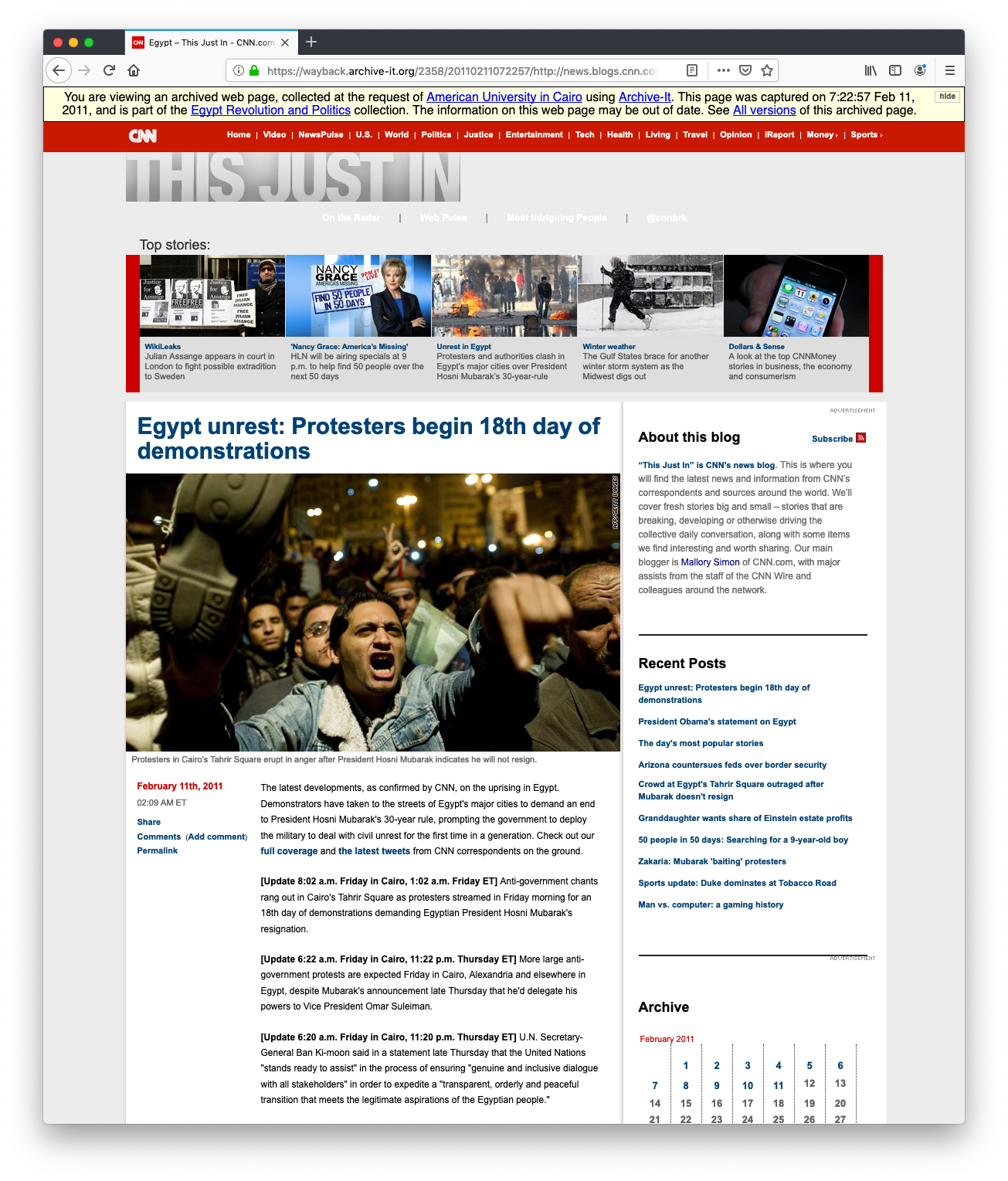}
\caption{This is a screenshot of the example CNN memento used in section \ref{sec:card_services}. Its memento-datetime is February 11, 2011, at 7:22:57 GMT and Archive-It has preserved it. This page was selected because it contains substantial content, including images.\\ URI: \url{https://wayback.archive-it.org/2358/20110211072257/http://news.blogs.cnn.com/category/world/egypt-world-latest-news/}}
\label{fig:cnn_example}
\end{figure}

The examples in this section will demonstrate some embeds using a memento of \emph{CNN Blogs} from 2011 preserved by Archive-It, shown in Figure \ref{fig:cnn_example}.

\begin{figure}
\centering
\includegraphics[width=0.6\textwidth]{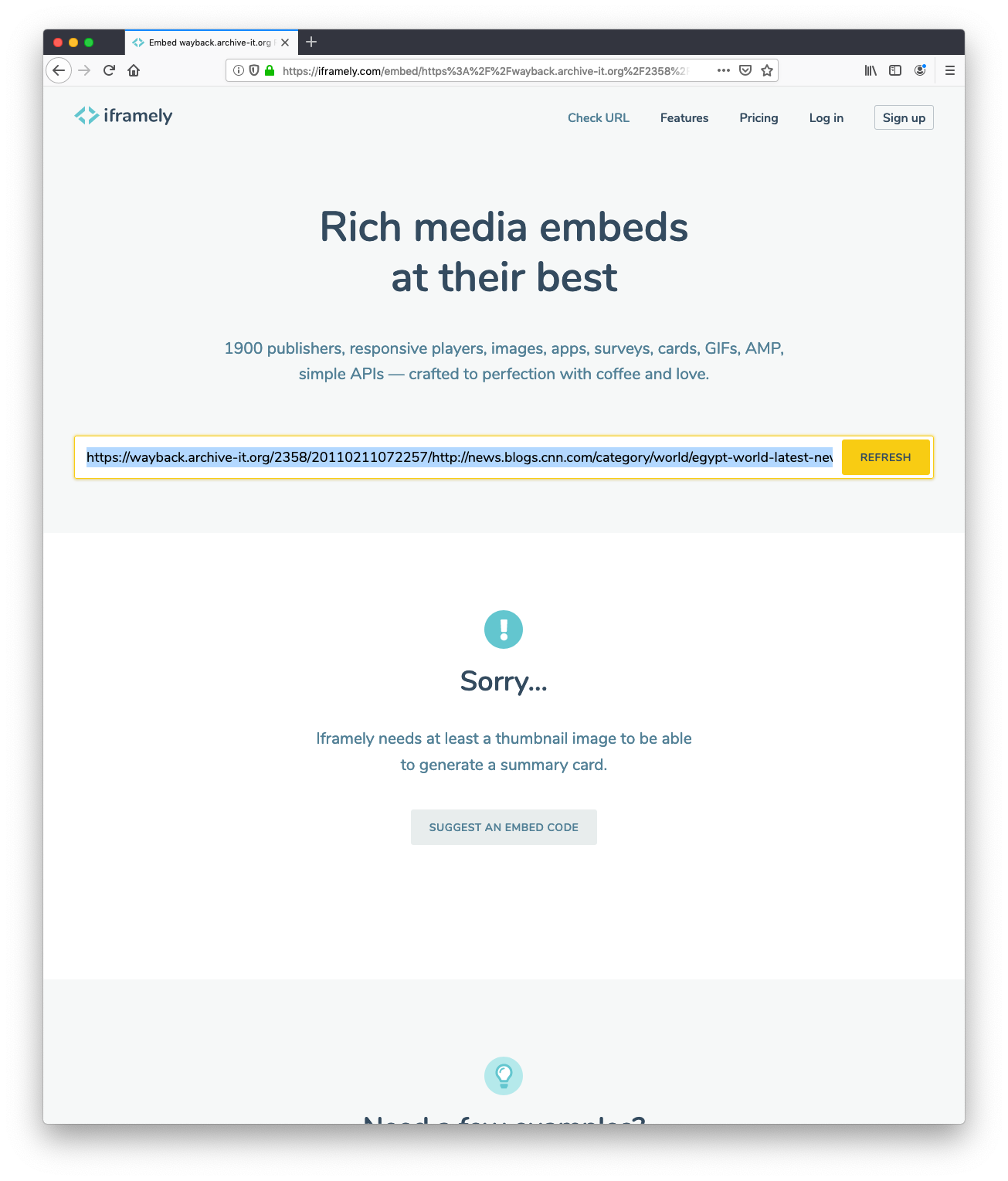}
\caption{What the Iframely parsers see for the Figure \ref{fig:cnn_example} according to their web application.}
\label{fig:iframely_error}
\end{figure}

In 2019, we reviewed Tumblr, Embed.ly, embed.rocks\footnote{\url{https://embed.rocks/}}, Iframely\footnote{\url{https://iframely.com/}}, noembed\footnote{\url{https://noembed.com/}}, microlink\footnote{\url{https://microlink.io/}}, and autoembed\footnote{\url{http://autoembed.com/}}. As of this writing, the autoembed service appears to be gone. The noembed service only provides embeds for a small number of web sites and does not support web archives. Iframely responds with errors for memento URIs, as shown in Figure \ref{fig:iframely_error}. Microlink produces output for some mementos, but not our CNN example.

\begin{figure}
\centering
\includegraphics[width=0.6\textwidth]{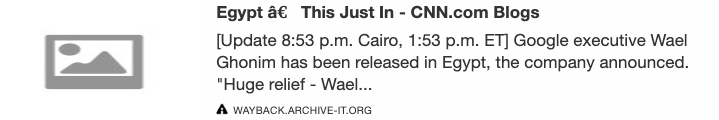}
\caption{The embed.rocks social card does not fare much better, attributing Figure \ref{fig:cnn_example}'s \emph{CNN} page to wayback.archive-it.org.}
\label{fig:bad_embedrocks}
\end{figure}

\begin{figure}[htbp]
  \centering
  \includegraphics[width=0.6\textwidth]{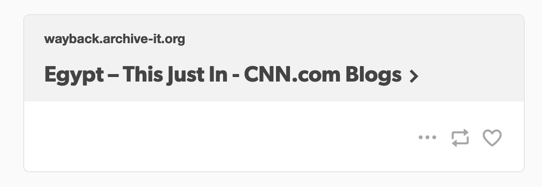}
  \caption{Tumblr cards misattribute content to web archives rather than their original source. Is this fake news?}
  \label{fig:tumblr_card}
\end{figure}

Tumblr, Embed.ly, and embed.rocks are the only services that produce output for our example memento. Unfortunately, none of these services are fully archive-aware. One of the goals of a good surrogate is to convey some level of aboutness for the underlying web resource. Mementos are documents with their own topics. They are typically not about the archives that contain them. Intermixing these two concepts of document content and archive information, without clear separation, produces surrogates that can confuse users. The embed.rocks service is not archive-aware. The embed.rocks example in Figure \ref{fig:bad_embedrocks} shows an embed that fails to convey the aboutness of its underlying memento, appearing to attribute this \emph{CNN} article to wayback.archive-it.org. Figure \ref{fig:tumblr_card} demonstrates that Tumblr has the same issues. What is the resource behind this surrogate really about? This mixing of resources weakens the surrogate's ability to convey the aboutness of the memento.

\begin{figure}
\centering
\includegraphics[width=0.6\textwidth]{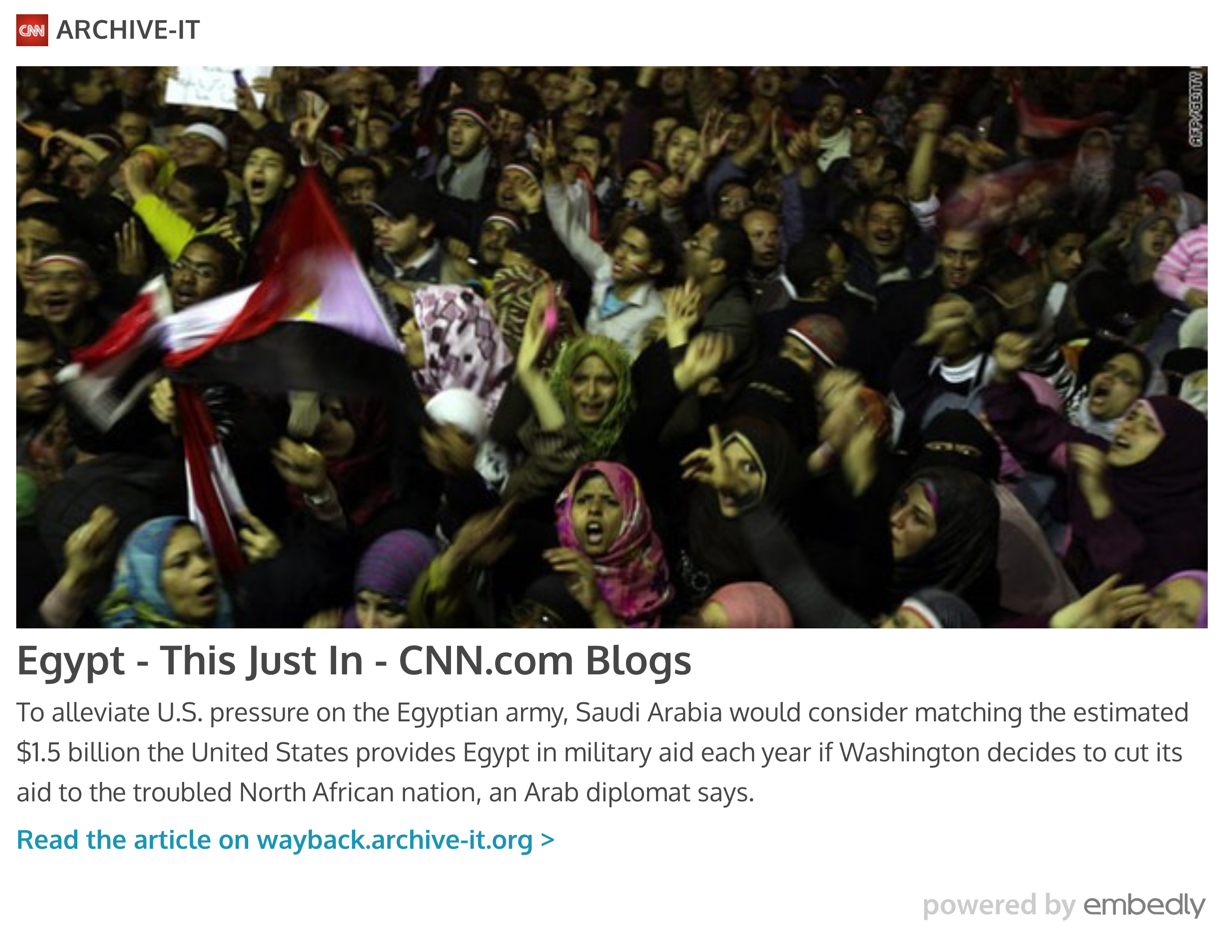}
\caption{This screenshot of an embed for the Figure \ref{fig:cnn_example}'s CNN memento shows how well Embed.ly performs. While the image and description convey more aboutness for the original resource, the memento is attributed to Archive-It with \emph{CNN's} favicon. Did \emph{CNN} or \emph{Archive-It} create this content? Should the reader be suspicious?}
\label{fig:bad_embedly}
\end{figure}

Embed.ly produces a much more attractive card but still falls short. In Figure \ref{fig:bad_embedly} we see an embed created for the same resource. It contains the title of the resource as well as a short description and even a striking image from the memento itself. Unfortunately, it contains no information about the original resource, potentially implying that someone at archive.org is serving content for CNN. Even worse, in the world where readers are concerned about fake news, this surrogate may lead an informed reader to believe that this is a link to a counterfeit resource because it does not come from cnn.com.

\clearpage

\section{Creating Surrogates With MementoEmbed}

\begin{figure}[t]
  \centering
  \includegraphics[width=0.7\textwidth]{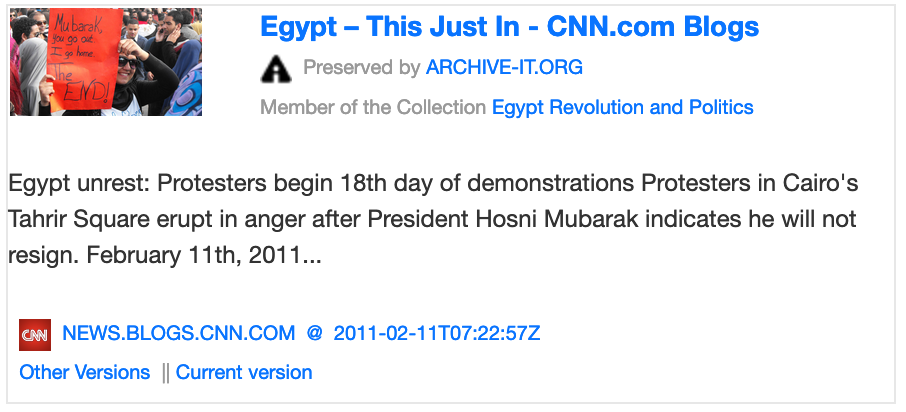}
  \caption{A social card generated by MementoEmbed based on the memento in Figure \ref{fig:cnn_example}.}
  \label{fig:socialcard-example}
\end{figure}

\begin{figure}
  \centering
  \includegraphics[width=0.9\textwidth]{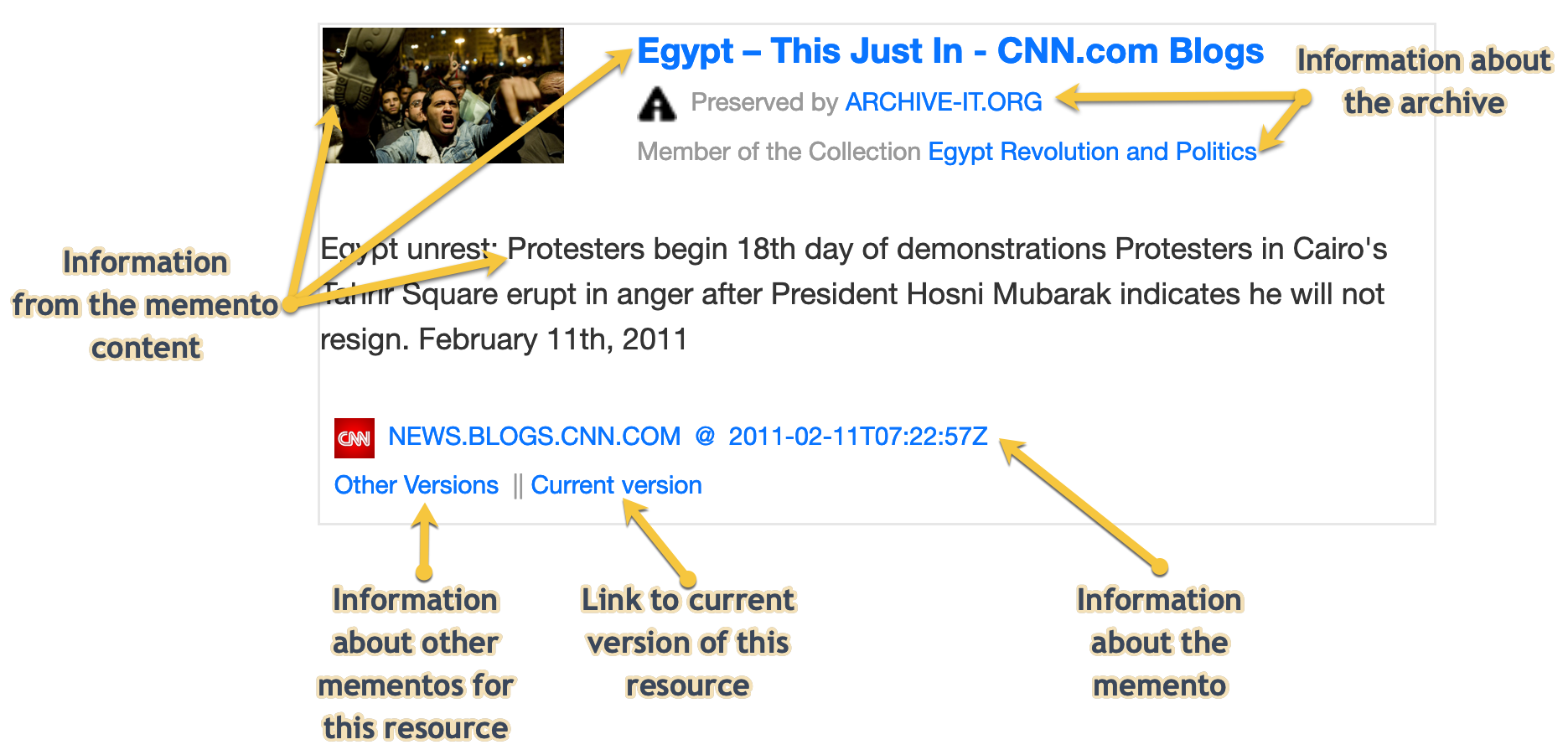}
  \caption{Because it is \emph{archive aware}, the MementoEmbed social card avoids the misattribution caused by other services.}
  \label{fig:mementoembed_annotated}
\end{figure}

Because of the issues detailed in the last section with existing platforms, we developed MementoEmbed to produce surrogates for mementos. MementoEmbed is the first \textbf{archive-aware} embeddable surrogate service, meaning it can include memento-specific information such as the memento-datetime, the archive from which a memento originates, and the memento's original resource domain name. Figure \ref{fig:socialcard-example} displays a social card generated by MementoEmbed. As demonstrated in Figure \ref{fig:mementoembed_annotated}, this card separates information derived from the memento content from information about the archive containing it. Through the Memento Protocol, MementoEmbed can attribute this content to its original web site, avoiding the confusion produced by other services, further allowing a user to understand the nature of the web page behind the surrogate.

\begin{figure}[t]
  \centering
\begin{subfigure}[t]{0.45\textwidth}
  % \centering
  \includegraphics[width=0.95\textwidth]{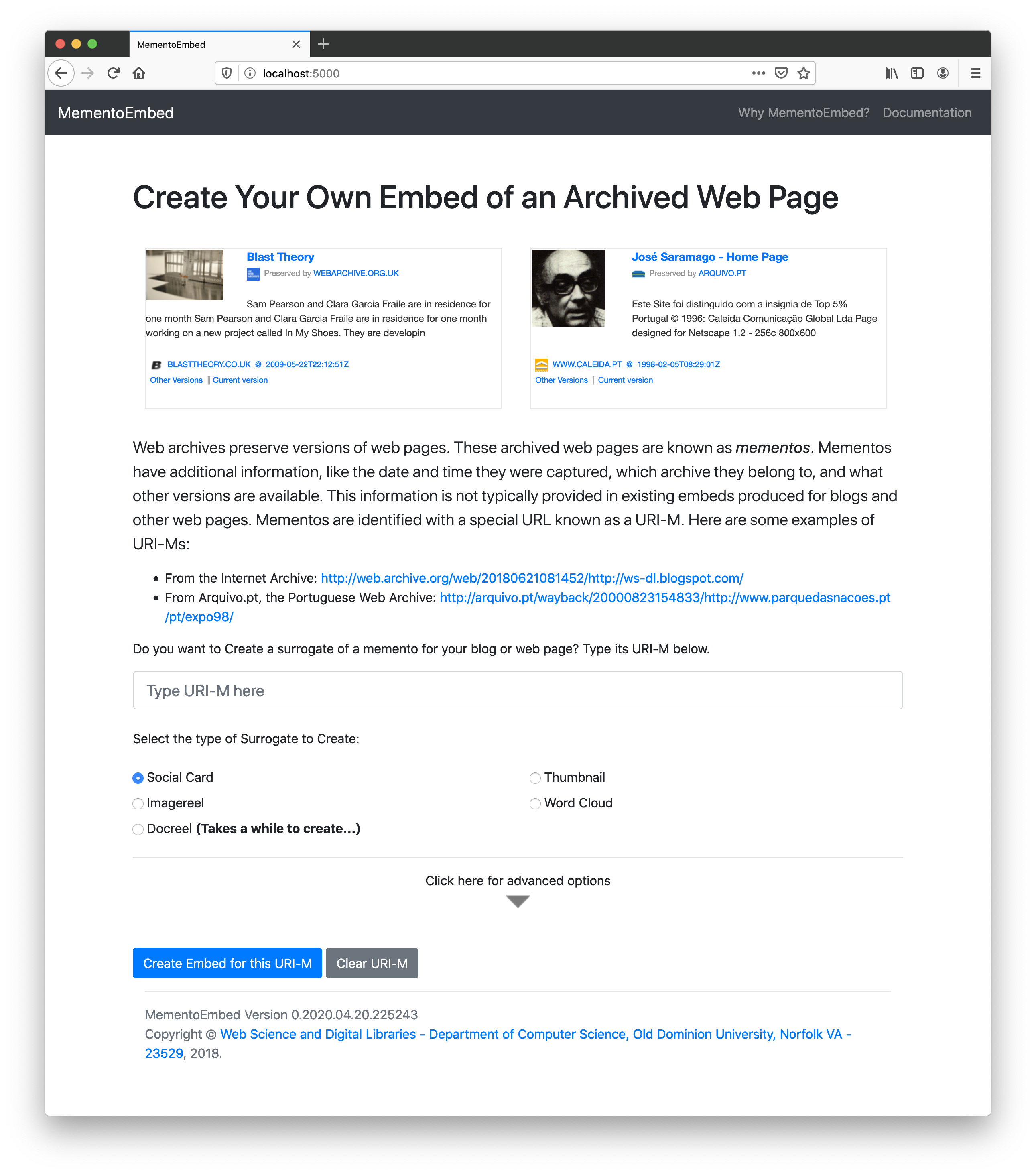}
  \caption{A user inserts a URI-M and selects their desired surrogate.}
  \label{fig:mementoembed_gui_start}
\end{subfigure}%
~
\begin{subfigure}[t]{0.45\textwidth}
  % \centering
  \includegraphics[width=\textwidth]{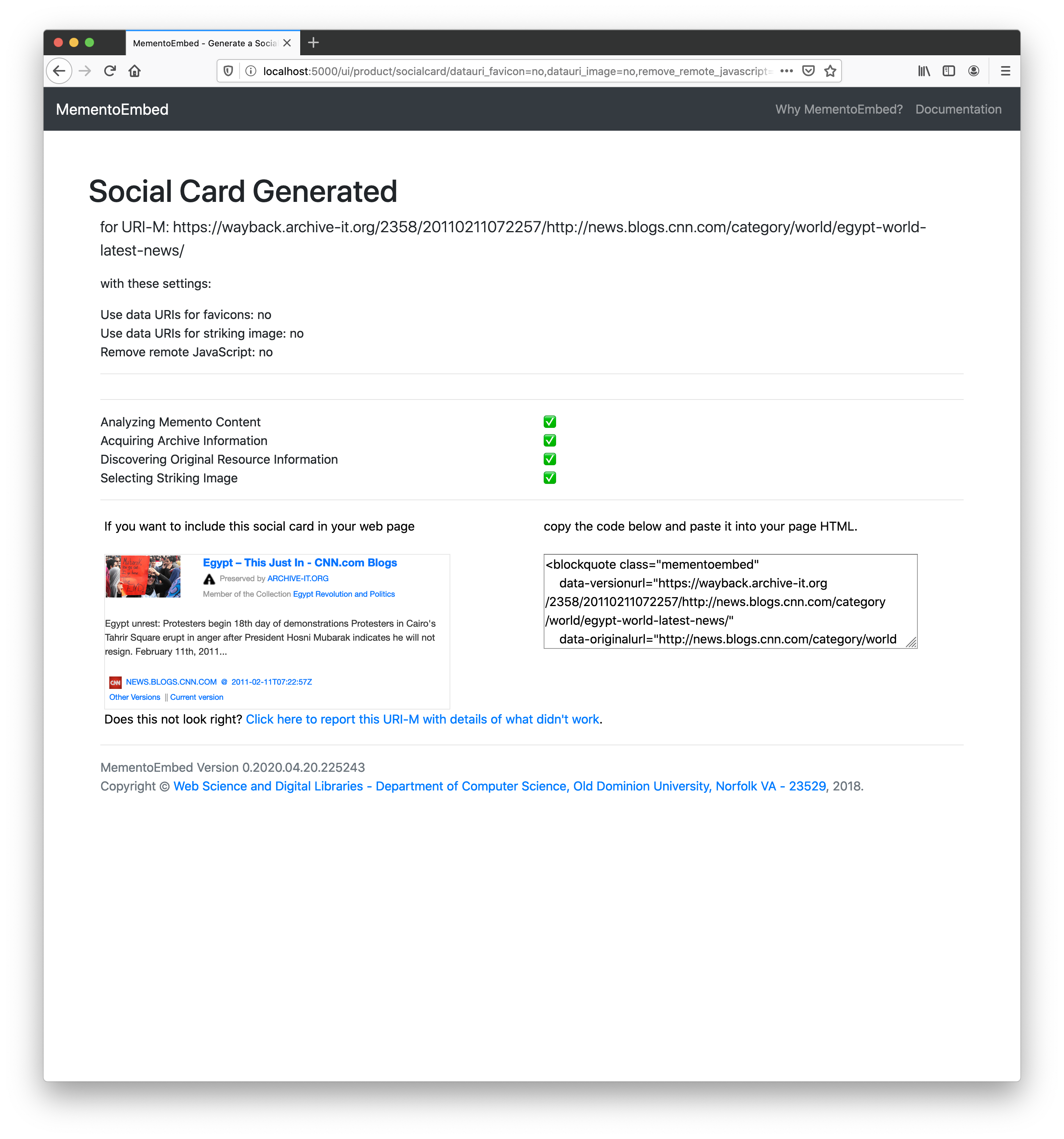}
  \caption{MementoEmbed displays the surrogate on the left and its associated HTML on the right.}
  \label{fig:mementoembed_gui_response}
\end{subfigure}

\caption{The MementoEmbed User Interface.}
\label{fig:mementoembed_gui}
\end{figure}

Figure \ref{fig:mementoembed_gui} demonstrates how a user can create their own cards. In Figure \ref{fig:mementoembed_gui_start}, the user enters the memento's URI-M and selects their desired surrogate type. MementoEmbed responds with a surrogate as shown in Figure \ref{fig:mementoembed_gui_response}. In addition to creating the surrogate, MementoEmbed also provides the user with code they can copy to embed the surrogate into their web page or blog.

\subsection{MementoEmbed Surrogate Types}

\begin{figure}[htbp]
  \centering
  \includegraphics[width=0.4\textwidth]{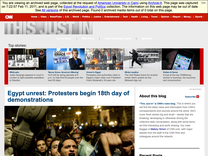}
  \caption{A browser thumbnail generated by MementoEmbed based on the memento in Figure \ref{fig:cnn_example}.}
  \label{fig:thumbnail_example}
% \end{figure}
\par\bigskip
% \begin{figure}[t]
\begin{subfigure}[t]{0.3\textwidth}
  \centering
  \includegraphics[width=\textwidth]{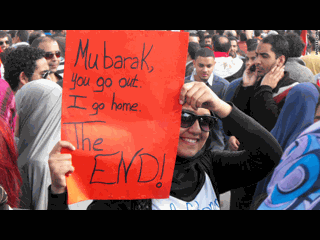}
\end{subfigure}%
~
\begin{subfigure}[t]{0.3\textwidth}
    \centering
    \includegraphics[width=\textwidth]{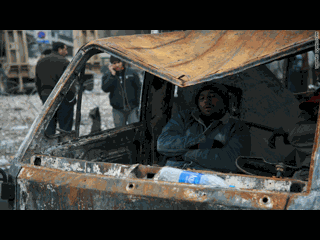}
\end{subfigure}
~
\begin{subfigure}[t]{0.3\textwidth}
    \centering
    \includegraphics[width=\textwidth]{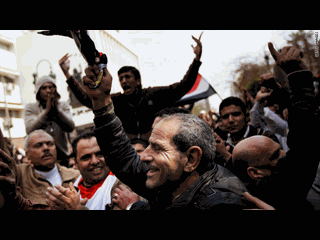}
\end{subfigure}

\par\smallskip

\begin{subfigure}[t]{0.3\textwidth}
  \centering
  \includegraphics[width=\textwidth]{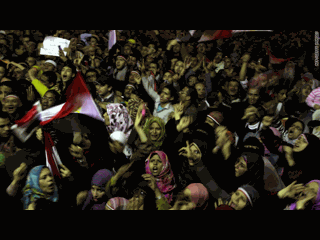}
\end{subfigure}%
~
\begin{subfigure}[t]{0.3\textwidth}
  \centering
  \includegraphics[width=\textwidth]{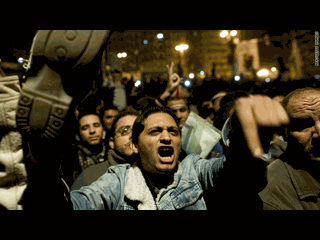}
\end{subfigure}%

\caption{The five images that make up MementoEmbed's imagereel created from Figure \ref{fig:cnn_example}'s memento. The imagereel starts black and fades to an image before returning back to black and fading to another image. The full animated GIF contains 105 frames.}
\label{fig:imagereel_example}

\par\bigskip

  \centering
  \includegraphics[width=0.4\textwidth]{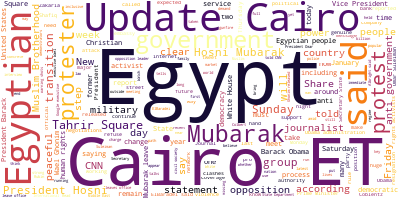}
  \caption{A word cloud generated by MementoEmbed based on the memento in Figure \ref{fig:cnn_example}.}
  \label{fig:wordcloud_example}
\end{figure}

\begin{figure}[htbp]

  \centering

  \begin{subfigure}[t]{0.3\textwidth}
    \centering
    \includegraphics[width=\textwidth]{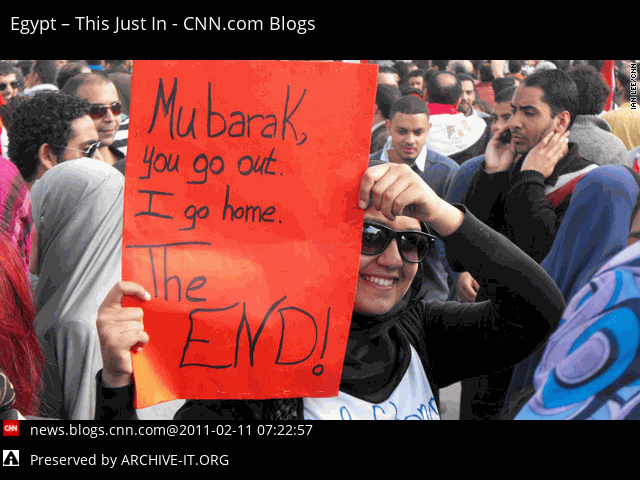}
  \end{subfigure}%
  ~
  \begin{subfigure}[t]{0.3\textwidth}
      \centering
      \includegraphics[width=\textwidth]{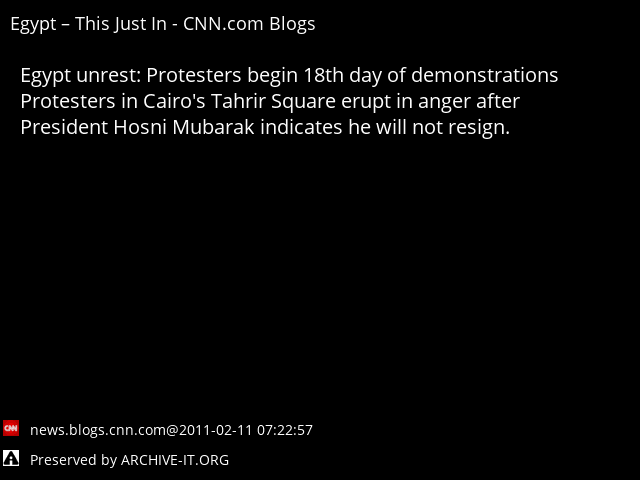}
  \end{subfigure}

  \begin{subfigure}[t]{0.3\textwidth}
      \centering
      \includegraphics[width=\textwidth]{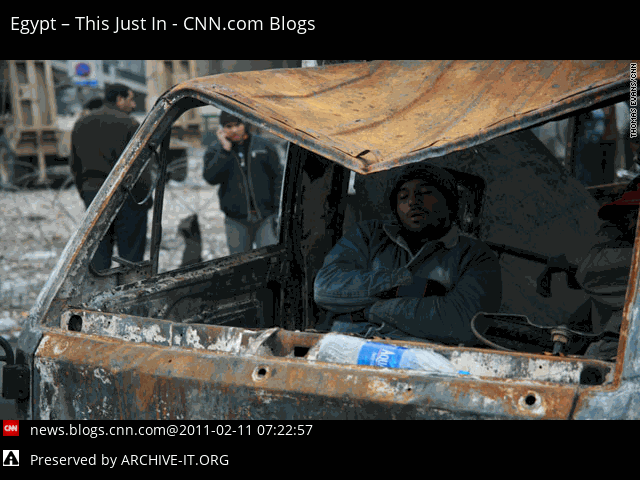}
  \end{subfigure}%
  ~
  \begin{subfigure}[t]{0.3\textwidth}
    \centering
    \includegraphics[width=\textwidth]{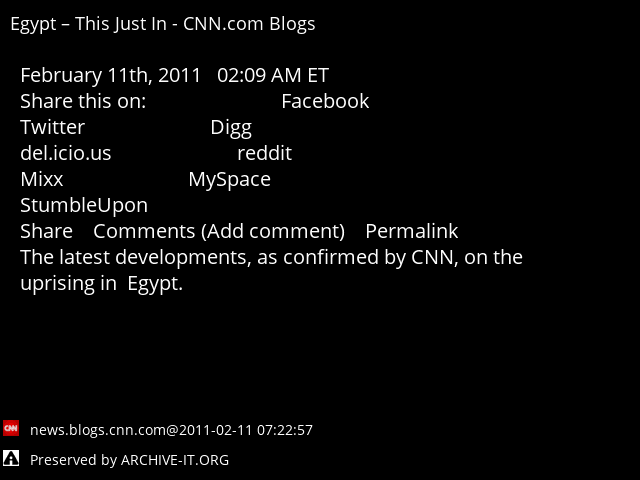}
  \end{subfigure}
  
  \begin{subfigure}[t]{0.3\textwidth}
    \centering
    \includegraphics[width=\textwidth]{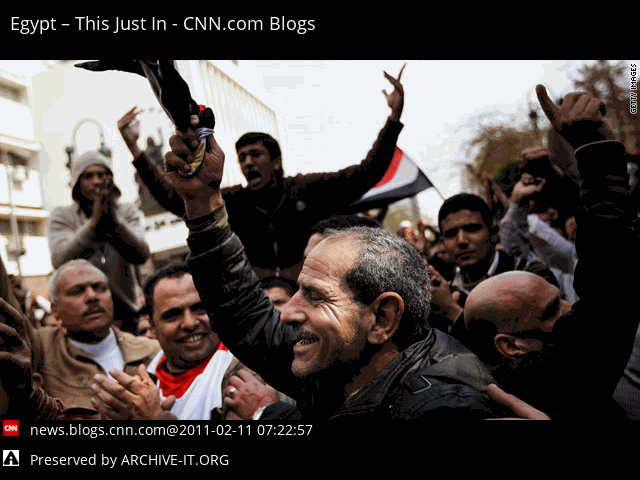}
  \end{subfigure}%
  ~
  \begin{subfigure}[t]{0.3\textwidth}
    \centering
    \includegraphics[width=\textwidth]{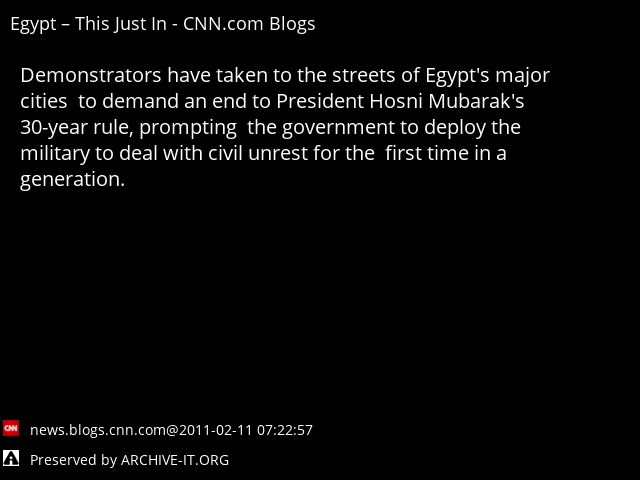}
  \end{subfigure}

  \begin{subfigure}[t]{0.3\textwidth}
    \centering
    \includegraphics[width=\textwidth]{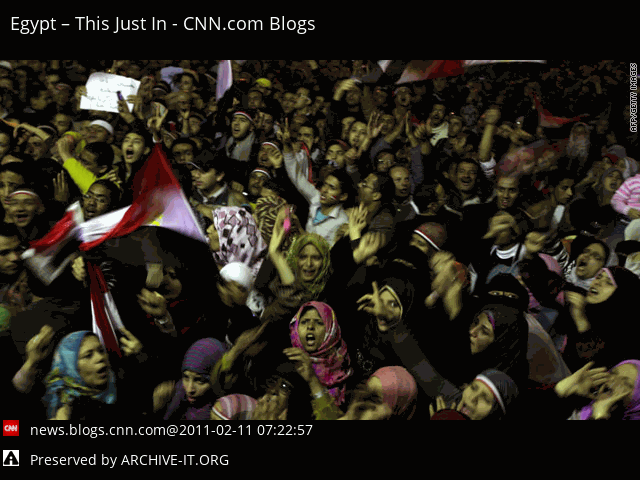}
  \end{subfigure}%
  ~
  \begin{subfigure}[t]{0.3\textwidth}
    \centering
    \includegraphics[width=\textwidth]{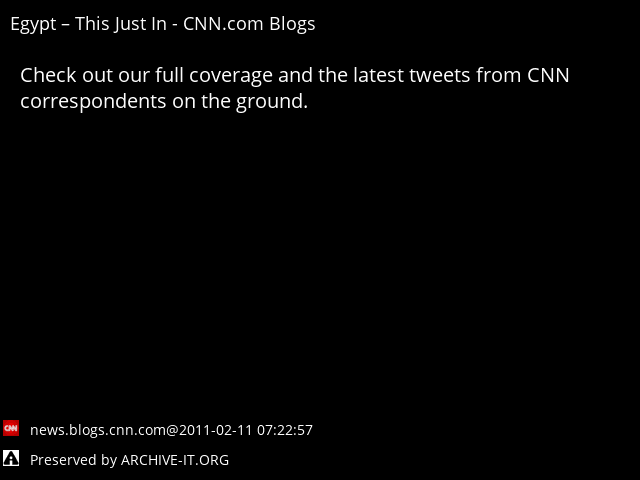}
  \end{subfigure}

  \begin{subfigure}[t]{0.3\textwidth}
    \centering
    \includegraphics[width=\textwidth]{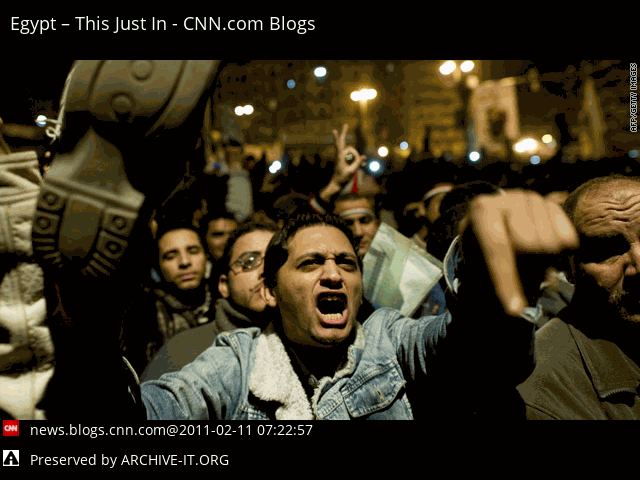}
  \end{subfigure}%
  ~
  \begin{subfigure}[t]{0.3\textwidth}
    \centering
    \includegraphics[width=\textwidth]{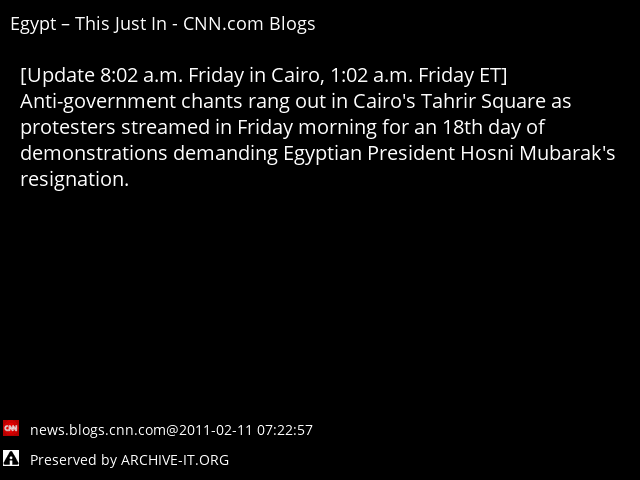}
  \end{subfigure}

\caption{The five images and five sentences that make up MementoEmbed's docreel created from Figure \ref{fig:cnn_example}'s memento. The docreel starts black and fades in an image or sentence before fading back to black.}
\label{fig:docreel_example}
\end{figure}

MementoEmbed currently generates the following surrogate types:
\begin{itemize}
  \item social cards (Figure \ref{fig:socialcard-example})
  \item browser thumbnails (Figure \ref{fig:thumbnail_example})
  \item \textbf{imagereels} (Figure \ref{fig:imagereel_example}) -- animated GIFs of the top five images
  \item word clouds (Figure \ref{fig:wordcloud_example})
  \item \textbf{docreels} (Figure \ref{fig:docreel_example}) --  animated GIFs of the top five images and top ranked sentences.
\end{itemize}

Working with web archives, we are acutely aware of how resources can disappear over time. We acknowledge that a user's web page may outlive the MementoEmbed instance from which they generated their surrogate. For this reason, the embeddable HTML code contains data URIs for all surrogate types but social cards, removing the requirement that the MementoEmbed service host images for other web pages.

MementoEmbed's documentation \cite{mementoembed_docs} details how to install and configure it. The following sections provide  detail on the available surrogates and how MementoEmbed generates them.

\subsubsection{Social Card}

MementoEmbed social cards, as seen in Figure \ref{fig:socialcard-example}, contain various pieces of information intended to summarize the resource:

\begin{itemize}
\item title
\item description
\item striking image
\item containing archive
\item archive favicon
\item original resource domain
\item original resource favicon
\item memento-datetime
\item a link to the original resource, if still present
\item a link to help readers discover other mementos of this resource
\end{itemize}

MementoEmbed uses the Memento Protocol to detect if the given URI-M identifies a memento. If so, it uses its knowledge of web archive URI patterns to discover the raw memento for the given URI-M. It then processes this raw memento to discover its title and description.

To discover a memento's title, MementoEmbed does the following with the raw memento's content:
\begin{enumerate}
  \item search for HTML \texttt{META} elements containing any of the following keys and save their values:
  \begin{itemize}
    \item \texttt{og:title}
    \item \texttt{twitter:title}
  \end{itemize}
  \item if a value for \texttt{og:title} is present, return that value
  \item if a value for \texttt{twitter:title} is present, return that value
  \item if no title has been found from this existing social media metadata, extract the text from the HTML \texttt{TITLE} element on the page and return it
  \item if no title has been found, return an empty string
\end{enumerate}

To discover a memento's description, MementoEmbed does the following with the raw memento's content:
\begin{enumerate}
  \item search for HTML \texttt{META} elements containing any of the following keys and save their values:
  \begin{itemize}
    \item \texttt{og:description}
    \item \texttt{twitter:description}
  \end{itemize}
  \item if a value for \texttt{og:description} is present, return that value
  \item if a value for \texttt{twitter:description} is present, return that value
  \item if no description has been found from this existing social media card metadata:
  \begin{enumerate}
    \item remove boilerplate from the HTML
    \item return the first 197 characters of the content, followed by an ellipsis if the last character is not punctuation
  \end{enumerate}
  \item if no description has been found, return an empty string
\end{enumerate}

To discover a memento's striking image, MementoEmbed uses the augmented memento's content rather than the raw memento's content. We want our card to contain an image from the web archive and not the live web. The augmented memento has rewritten URIs for its images and will ensure that MementoEmbed only processes images that have been archived. It uses the following process to discover images:
\begin{enumerate}
  \item search for HTML \texttt{META} elements containing any of the following keys and save their values:
  \begin{itemize}
    \item \texttt{og:image}
    \item \texttt{twitter:image}
    \item \texttt{twitter:image:src}
    \item \texttt{image}
  \end{itemize}
  \item if a value for \texttt{og:image} is present, is a memento, and its URI-M returns a 200 HTTP status, return that value
  \item if a value for \texttt{twitter:image} is present, is a memento, and its URI-M returns a 200 HTTP status, return that value
  \item if a value for \texttt{twitter:image:src} is present, is a memento, and its URI-M returns a 200 HTTP status, return that value
  \item if a value for \texttt{image} is present, is a memento, and its URI-M returns a 200 HTTP status, return that value
  \item if no image URI has been found from this existing social media metadata:
  \begin{enumerate}
    \item extract all image URIs from the \texttt{SRC} attribute of \texttt{IMG} HTML elements
    \item extract all image URIs from the \texttt{SRCSET} attribute of \texttt{IMG} HTML elements
    \item for each image URI:
    \begin{enumerate}
      \item if URI does not identify a memento, use datetime negotation to discover a memento of the image, if possible
      \item download the corresponding image and calculate terms for the scoring equation, as described below
    \end{enumerate}
    \item select the highest scoring image
  \end{enumerate}
  \item if no image has been found, return the default image of a wireframe globe
\end{enumerate}

The goal of image selection is to discover an image quickly. Thus, MementoEmbed does not use external services like Google's Vision API\footnote{\url{https://cloud.google.com/vision/}} or Amazon Rekognition API\footnote{\url{https://docs.aws.amazon.com/rekognition/latest/dg/images.html}} to analyze images. With no pre-existing knowledge of the memento, the system attempts to use easily discoverable image features to score each image with Equation \ref{eq:mementoembed_scoring}

\begin{equation}
  S = k_1 (N - n) + (k_2 s) - (k_3 h) - (k_4 r) + (k_5 c) 
  % k_1 = 0.1 \\
  % k_2 = 0.4 \\
  % k_3 = 10 \\
  % k_4 = 0.5 \\
  % k_5 = 10 \\
  \label{eq:mementoembed_scoring}
\end{equation}

\noindent where $N$ is the number of images in the HTML, $n$ is the image position as discovered in the HTML, $s$ is the image size in pixels as width $\times$ height, $h$ is the number of columns in the image histogram where the value is 0, $r$ is the ratio of image width to height, and $c$ is the number of colors in the image. The terms $k_1$ through $k_5$ are weights applied to these features' raw values. Through trial and error, we have discovered that the most desirable images occur earlier in the page (low $n$), are larger (high $s$), have less negative space (low $h$), are less likely to look like a rectangular banner (low $r$), and are colorful (high $c$). Thus, we optimized the equation for these attributes.

MementoEmbed generates the containing archive name by extracting the registered domain name from the URI-M and making it upper case. We decided it best to present the domain name to users rather than matching domain names to organizational names for a few reasons. Organizations have different names in different languages. Organizations may change their names over time. The registered domain name is a sufficient form of identification, promotes the archive, and allows the reader to discover more about the archive through this link.

For additional archive branding, we discover the web archive's favicon and apply it to the card. To discover this, MementoEmbed employs the following algorithm:
\begin{enumerate}
  \item extract the domain name and scheme from the URI-M to discover the web archive's home page
  \item visit this home page
  \item analyze the HTML in the home page to find the favicon in the \texttt{LINK} element's \texttt{icon}, \texttt{shortcut}, or \texttt{shortcut icon} relations -- return this value if present
  \item if no valid favicon, construct the URI of the favicon by appending \texttt{favicon.ico} to the web archive's home page URI -- if this URI resolves, return this value
  \item if no valid favicon thus far, then consult the Google Favicon service\footnote{\url{https://www.google.com/s2/favicons?domain=[INSERT URI]}} using the web archive's domain name as the input parameter
\end{enumerate}

While downloading a memento, MementoEmbed extracts the URI-R and memento-datetime from the headers provided by the Memento Protocol. It applies the URI-R and memento-datetime to the card. It parses the URI-R to discover the original resource domain name. We discover the original resource favicon with the following algorithm:
\begin{enumerate}
  \item analyze the HTML in the memento to find the favicon in the \texttt{LINK} element's \texttt{icon}, \texttt{shortcut}, or \texttt{shortcut icon} relations 
  \item if this value is present
  \begin{enumerate}
    \item verify that it is a memento, if so, return it
    \item if not, use datetime negotiation to discover a memento of this URI
    \item if a memento of this URI is present, return its URI-M
  \end{enumerate}
  \item if MementoEmbed has not discovered a favicon thus far, construct the URI of the favicon by appending \texttt{favicon.ico} to the orignal resource domain URI and perform datetime negotiation to find its URI-M -- if this URI-M resolves, return 
  \item if no URI-M for this favicon, then test that the URI-R exists, and if so, return that URI-R if it returns an HTTP 200 status code
  \item if MementoEmbed has not discovered a favicon, then consult the Google Favicon service\footnote{\url{https://www.google.com/s2/favicons?domain=[INSERT URI]}} using the original resource's domain name as the input parameter
\end{enumerate}

MementoEmbed provides a link to other mementos through the Los Alamos National Laboratory's Memento Aggregator Time Travel service\footnote{\url{http://timetravel.mementoweb.org/}}. To build the link, MementoEmbed appends the memento datetime and the URI-R to the Aggregator's API -- e.g., for the memento in Figure \ref{fig:cnn_example} it generates \url{http://timetravel.mementoweb.org/list/20110211072257Z/http://news.blogs.cnn.com/category/world/egypt-world-latest-news/}.

Users can control the creation of this social card with several advanced options. They can convert the striking image and favicons to data URIs \cite{masinter_rfc_1998}, thus ensuring that their card will be unaffected by issues with the web archive containing these mementos. By default, MementoEmbed cards require a JavaScript library that is hosted with the MementoEmbed instance. To avoid dependencies on the MementoEmbed service, the user can select an option that will generate a card without any dependency on this JavaScript library.

\subsubsection{Browser Thumbnail}

To produce a browser thumbnail as shown in Figure \ref{fig:thumbnail_example}, MementoEmbed uses Puppeteer\footnote{\url{https://developers.google.com/web/tools/puppeteer}} to load the URI-M in a headless version of Google Chrome and capture a screenshot. Users have the option to change the height and width of the thumbnail. They can also alter the viewport size of the headless browser. They can specify a timeout for MementoEmbed to stop trying if the thumbnail takes an excessive amount of time to generate. They can also ask MementoEmbed to remove a web archive's banner from the thumbnail. This last option is only available for web archives that support separate URI-Ms that do not contain this banner.

\subsubsection{Imagereel}

MementoEmbed can produce a new type of surrogate we refer to as the \textbf{imagereel}. Imagereels are animated GIFs containing a fixed number of images extracted from the underlying memento. MementoEmbed orders the images by their descending Equation \ref{eq:mementoembed_scoring} score. MementoEmbed then applies a fading effect to each image, creating frames where the image fades from black to the full image and back to black again. MementoEmbed then arranges these frames so that the resulting imagereel displays each image fading in and out. As print documents cannot display animated GIFs, in Figure \ref{fig:imagereel_example}, we instead display the non-faded frames of the imagereel produced from the memento from Figure \ref{fig:cnn_example}. 

The default number of images is five. A user can change the width and height of the imagereel, the duration of each frame, and the number of images to include. Imagereels are only really practical for mementos that contain many images, thus limiting their use. They can be combined with other content in a blog post to feature a variety of images in a smaller space, similar to the carousel \cite{pernice_2013} present on many modern web pages. They can also be used in a social media post to provide a large number of images in a small space.

\subsubsection{Word Cloud}

Word clouds \cite{10.1145/1374489.1374501} are commonly used to convey the meaning of documents by displaying words in sizes proportional to their frequency. MementoEmbed leverages the wordcloud \cite{wordcloud_python} Python library to produce word clouds from the memento, as shown in Figure \ref{fig:wordcloud_example}. Word clouds are a newer addition to MementoEmbed. They are not yet configurable via the graphical user interface but are configurable via the API covered in Section \ref{sec:mementoembed_webapi}.

\subsubsection{Docreel}

MementoEmbed also includes a second new type of surrogate we refer to as the \textbf{docreel}. Like imagereels, docreels are animated GIFs. In addition to the top-scoring images in the memento, docreels include the memento's top-scoring sentences. The images are scored using Equation \ref{eq:mementoembed_scoring} and the sentences are scored by the ARC90 readability \cite{andreasvc_python} algorithm. In addition to the images and sentences, MementoEmbed overlays the original domain names, archive domain names, favicons, and the memento-datetime onto each frame.  Again, as with imagreels, the images and sentences fade in and out. Figure \ref{fig:docreel_example} displays ten non-faded frames from an example docreel. Docreels are experimental, can take more than five minutes to generate, and the browser may timeout before MementoEmbed completes their generation. They are not currently configurable by users.

\subsection{Web API}
\label{sec:mementoembed_webapi}

We developed MementoEmbed's API in concert with its user interface. Originally, our goal with the API was to allow the user interface to make multiple simultaneous requests to improve response times. It has evolved into a more general-purpose memento information and visualization tool. This section will only provide an overview of the API. For more detailed and current information consult the documentation \cite{mementoembed_api_docs}. The two main branches of the MementoEmbed web API are:

\begin{itemize}
\item \texttt{/services/memento} - for getting specific information about a memento, detailed in Table \ref{tab:services_memento_fields} in Appendix A
\item \texttt{/services/product} - for directly producing a surrogate through one of the following:
\begin{itemize}
  \item  \texttt{/services/product/socialcard/}
  \item  \texttt{/services/product/thumbnail/}
  \item  \texttt{/services/product/imagereel/}
  \item  \texttt{/services/product/wordcloud/}
  \item  \texttt{/services/product/docreel/}
\end{itemize}
\end{itemize}

Clients request information by choosing the appropriate endpoint and appending the URI-M to that endpoint. Each endpoint responds with a JSON object containing fields and data about that particular aspect of the memento. For example, Figure \ref{fig:contentdata_example_output} displays the JSON response for the URI-M \url{https://www.webarchive.org.uk/wayback/archive/20090522221251/http://blasttheory.co.uk/} submitted to the \texttt{/services/memento/contentdata/} endpoint. Every endpoint returns the fields \texttt{urim} and \texttt{generation-time} by default. The \texttt{contentdata} endpoint provides the additional fields of \texttt{title} and \texttt{snippet} drawn from the content of the memento. Other endpoints have their own fields for providing additional memento data, as listed in Table \ref{tab:services_memento_fields} of Appendix A.

\begin{figure}[t]
\begin{lstlisting}
{
  "urim": "https://www.webarchive.org.uk/wayback/archive/20090522221251/http://blasttheory.co.uk/",
  "generation-time": "2020-04-20T22:59:52Z",
  "title": "Blast Theory",
  "snippet": "Sam Pearson and Clara Garcia Fraile are in residence for one month Sam Pearson and Clara Garcia Fraile are in residence for one month working on a new project called In My Shoes. They are developin...",
  "memento-datetime": "2009-05-22T22:12:51Z"
}
\end{lstlisting}
\caption{Example output from MementoEmbed's \url{/services/memento/contentdata/https://www.webarchive.org.uk/wayback/archive/20090522221251/http://blasttheory.co.uk/} API endpoint}
\label{fig:contentdata_example_output}
\end{figure}

\begin{figure}[htbp]
\begin{lstlisting}
{
  "urim": "https://www.webarchive.org.uk/wayback/archive/20090522221251/http://blasttheory.co.uk/",
  "processed urim": "https://www.webarchive.org.uk/wayback/archive/20090522221251im_/http://blasttheory.co.uk/",
  "generation-time": "2020-04-20T23:01:00Z",
  "images": {
      "https://www.webarchive.org.uk/wayback/archive/20090522221251im_/http:/blasttheory.co.uk/bt/i/uncleroy/ur_icon.jpg": {
          "source": "body",
          "content-type": "image/jpeg",
          "is-a-memento": true,
          "magic type": "JPEG image data, JFIF standard 1.02, aspect ratio, density 100x100, segment length 16, progressive, precision 8, 168x96, components 3",
          "imghdr type": "jpeg",
          "format": "JPEG",
          "mode": "RGB",
          "width": 168,
          "height": 96,
          "blank columns in histogram": 463,
          "size in pixels": 16128,
          "ratio width/height": 1.75,
          "byte size": 1979,
          "colorcount": 4776,
          "pHash": "f771185c8c86c667",
          "aHash": "00008085f7ffffff",
          "dHash_horizontal": "0c1b4b0d0d2c2c4d",
          "dHash_vertical": "cbf7ffffffffffff",
          "wHash": "00008080e7ffffff",
          "N": 14,
          "n": 11,
          "k1": 0.1,
          "k2": 0.4,
          "k3": 10,
          "k4": 0.5,
          "k5": 10,
          "calculated score": 49580.625
      },
      ... other records omitted for brevity ...
  },
  "ranked images": [
      "https://www.webarchive.org.uk/wayback/archive/20090522221251im_/http:/blasttheory.co.uk/bt/i/dotf/Untitled-1.jpg",
      "https://www.webarchive.org.uk/wayback/archive/20090522221251im_/http:/blasttheory.co.uk/bt/i/yougetme/ygm_icon.jpg",
      "https://www.webarchive.org.uk/wayback/archive/20090522221251im_/http:/blasttheory.co.uk/bt/i/cysmn/cy_icon.jpg",
      "https://www.webarchive.org.uk/wayback/archive/20090522221251im_/http:/blasttheory.co.uk/bt/i/rider_spoke/rs_icon.jpg",
      "https://www.webarchive.org.uk/wayback/archive/20090522221251im_/http:/blasttheory.co.uk/bt/i/ulrikeandeamon/ulrikeandeamon_small.jpg",
      "https://www.webarchive.org.uk/wayback/archive/20090522221251im_/http:/blasttheory.co.uk/bt/i/trucold/trucold_icon.jpg",
      "https://www.webarchive.org.uk/wayback/archive/20090522221251im_/http:/blasttheory.co.uk/bt/i/uncleroy/ur_icon.jpg",
      "https://www.webarchive.org.uk/wayback/archive/20090522221251im_/http:/blasttheory.co.uk/bt/pe/bt_logo.gif",
      "https://www.webarchive.org.uk/wayback/archive/20090522221251im_/http:/blasttheory.co.uk/bt/pe/latest.gif",
      "https://www.webarchive.org.uk/wayback/archive/20090522221251im_/http:/blasttheory.co.uk/bt/pe/about.gif",
      "https://www.webarchive.org.uk/wayback/archive/20090522221251im_/http:/blasttheory.co.uk/bt/pe/home.gif",
      "https://www.webarchive.org.uk/wayback/archive/20090522221251im_/http:/blasttheory.co.uk/bt/pe/recent.gif",
      "https://www.webarchive.org.uk/wayback/archive/20090522221251im_/http:/blasttheory.co.uk/bt/pe/types.gif",
      "https://www.webarchive.org.uk/wayback/archive/20090522221251im_/http:/blasttheory.co.uk/bt/pe/chrono.gif"
  ]
}
\end{lstlisting}
\caption{Some example output from the \url{/services/memento/imagedata/https://www.webarchive.org.uk/wayback/archive/20090522221251/http://blasttheory.co.uk/} API endpoint}
\label{fig:imagedata_example_output}
\end{figure}

The MementoEmbed API uses the following HTTP status codes to indicate success or error:
\begin{itemize}
  \item 200 - all was successful
  \item 400 - the submitted URI is invalid, either lacking a scheme or otherwise not formed according to standards
  \item 404 - the submitted URI does not identify a memento, and MementoEmbed cannot service it
  \item 500 - any error not covered elsewhere
  \item 502 - while requesting the URI-M for processing, there was a connection issue with the web archive
  \item 504 - while requesting the URI-M from the archive for processing, the connection timed out
\end{itemize}

Because page content varies, some endpoints provide far more information than is detailed in Table \ref{tab:services_memento_fields} of Appendix A. For example, Figure \ref{fig:imagedata_example_output} shows the wealth of information available for images discovered in the memento, including content-type, pixel size, perceptive hash (pHash), details of the weights used in the scoring equation, and much more. Developers who wish to use the API should consult the full API documentation \cite{mementoembed_api_docs} for more details on what sub-fields are available for pages where the output fields vary based on the memento, such as the the \texttt{imagedata}, \texttt{seeddata}, \texttt{paragraphrank}, \texttt{setencerank}, and \texttt{page-metadata} endpoints.

Clients can alter the behavior of some endpoints via the \texttt{Prefer} \cite{snell_rfc_2014} HTTP request header. For example, the default width of thumbnails is 208 pixels with a viewport width of 1024 pixels. If a client desires a thumbnail 2048 pixels wide with a viewport of 4096 pixels wide, they specify these values in the \texttt{Prefer} header as shown in \ref{fig:thumbnail_prefer_request}. In Figure \ref{fig:thumbnail_prefer_response}, the server uses the \texttt{Preference-Applied} response header to list the values of all preferences used when generating the resulting thumbnail. Figure \ref{tab:preferences_api_services} provides a list of all endpoints and their available preferences.

The user interface is a client of the API. We did not employ specialized logic generate surrogates differently in the API compared to the user interface. We developed MementoEmbed with the intention that the API would be used by a variety of clients to gather information on individual mementos. Raintale is one such client. We developed Raintale for telling stories with many mementos.

\begin{figure}
  \begin{subfigure}[t]{\textwidth}
\begin{lstlisting}
GET /services/product/thumbnail/http://web.archive.org/web/20180128152127/http://www.cs.odu.edu/~mkelly/ HTTP/1.1
Host: mementoembed.ws-dl.cs.odu.edu
User-Agent: curl/7.54.0
Accept: */*
Prefer: viewport_width=4096,thumbnail_width=2048
\end{lstlisting}
\caption{The request specifies these options as part of the \texttt{Prefer} header.}
\label{fig:thumbnail_prefer_request}
\end{subfigure}
\par\bigskip
\begin{subfigure}[t]{\textwidth}
  \begin{lstlisting}
HTTP/1.0 200 OK
Content-Type: image/png
Content-Length: 437589
Preference-Applied: viewport_width=4096,viewport_height=768,thumbnail_width=2048,thumbnail_height=156,timeout=60
Server: Werkzeug/0.14.1 Python/3.6.5
Date: Wed, 25 Jul 2018 20:59:21 GMT

...437589 bytes of data follow...
  \end{lstlisting}
\caption{The response indicates which preferences were applied via the \texttt{Preference-Applied} header.}
\label{fig:thumbnail_prefer_response}
\end{subfigure}
\caption{HTTP request and response with the web API for a browser thumbnail with a viewport width of 4096 pixels and a thumbnail width of 2048 pixels.}
\end{figure}

\clearpage

\section{Telling Stories From Web Archives With Raintale}
\label{sec:raintale}

\begin{figure}[t]
  \centering
  \includegraphics[width=0.8\textwidth]{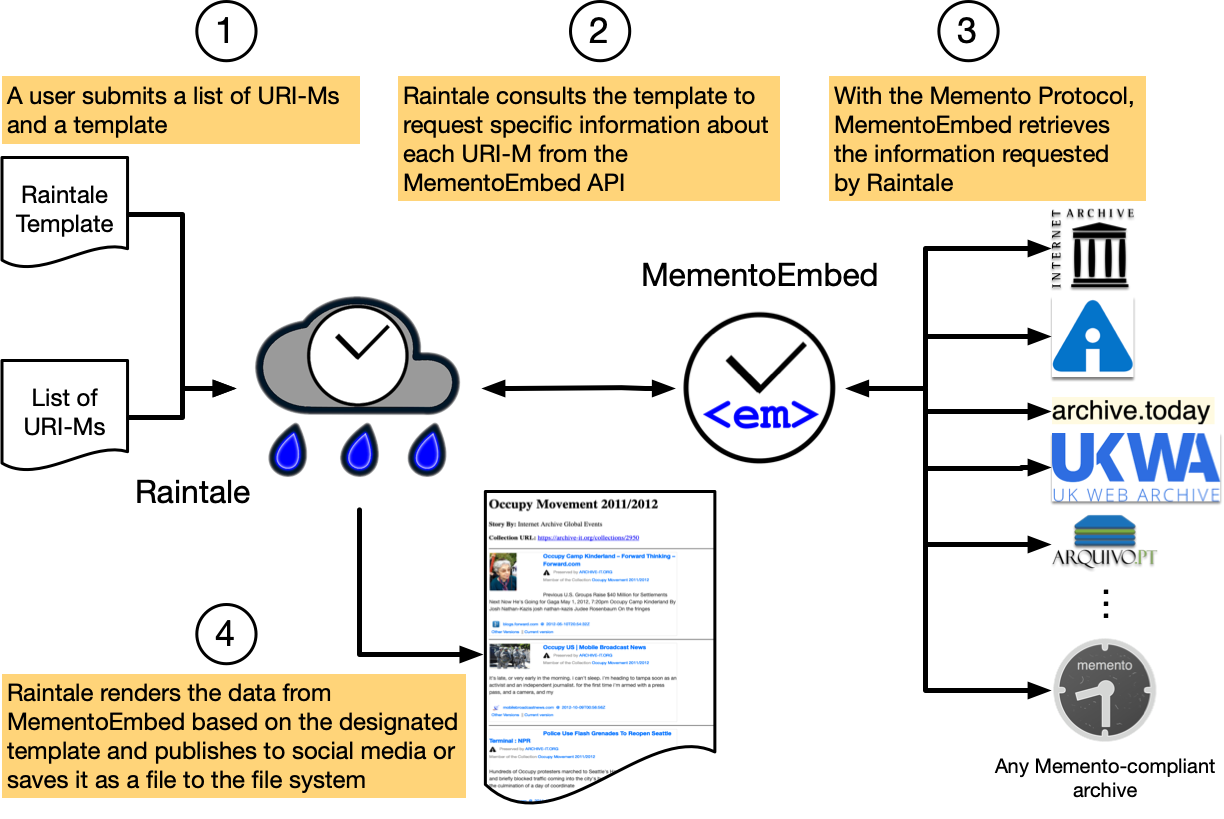}
  \caption{The MementoEmbed-Raintale Architecture for Storytelling}
  \label{fig:memento-embed-raintale-integration}
\end{figure}

Where MementoEmbed focuses on a single memento, Raintale provides stories created from multiple mementos. It accepts two forms of input: a list of URI-Ms and a template. The template formats the output and contains variables that indicate what information the user wishes to display for each memento. Figure \ref{fig:memento-embed-raintale-integration} demonstrates the relationship between MementoEmbed and Raintale. In step 1,  the user provides a template and a list of URI-Ms to Raintale. In step 2, Raintale records all template variables. For each provided URI-M, Raintale consults MementoEmbed's API for each variable's value in the corresponding memento. In step 3, MementoEmbed downloads the memento from its web archive and performs natural language processing, image analysis, or extracts information via the Memento Protocol \cite{van_de_sompel_rfc_2013}, as appropriate to the API request. In step 4, Raintale consolidates the data from these API responses and renders the template with the gathered data. It produces a story constructed from surrogates and other supplied content (e.g., story title, collection metadata).

Raintale is currently a command-line application that provides various options for the end user. To create a story using the mementos supplied in \texttt{story-mementos.txt} and a template file \texttt{mytemplate.tmpl} and save the output to \texttt{mystory.html}, a user would type the following.
\begin{lstlisting}[frame=none,postbreak=]
  # tellstory -i story-mementos.txt --storyteller template --story-template mytemplate.tmpl -o mystory.html --title "This is My Story Title"
\end{lstlisting}

We designed Raintale this way so that users can easily integrate it into shell scripts with other utilities, such as Hypercane \cite{jones_hypercane_2020_1,jones_hypercane_2020_2,jones_hypercane_2020_3}, that can provide it a list of mementos or an automated utility that might produce different templates for different use cases. We intend for Raintale to be part of an automated web archiving workflow, and we provide features that allow the archivist to customize its output to their needs.

\begin{figure}[H]
  \centering

  \includegraphics[width=0.5\textwidth]{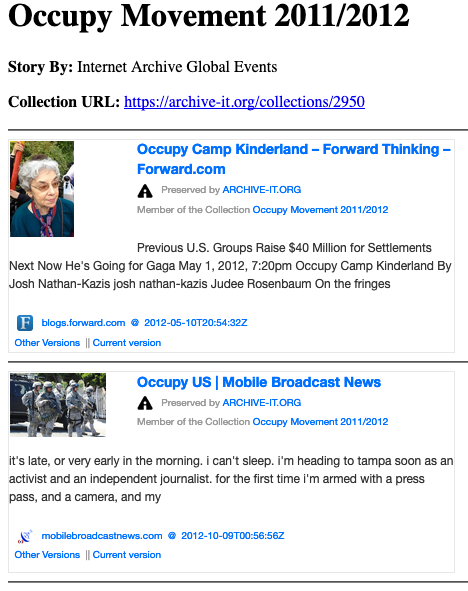}
  \caption{Raintale output examples - HTML with social cards}
  \label{fig:raintale-template-default-html}
\end{figure}

\begin{figure}[H]

  \centering
  \includegraphics[width=\textwidth]{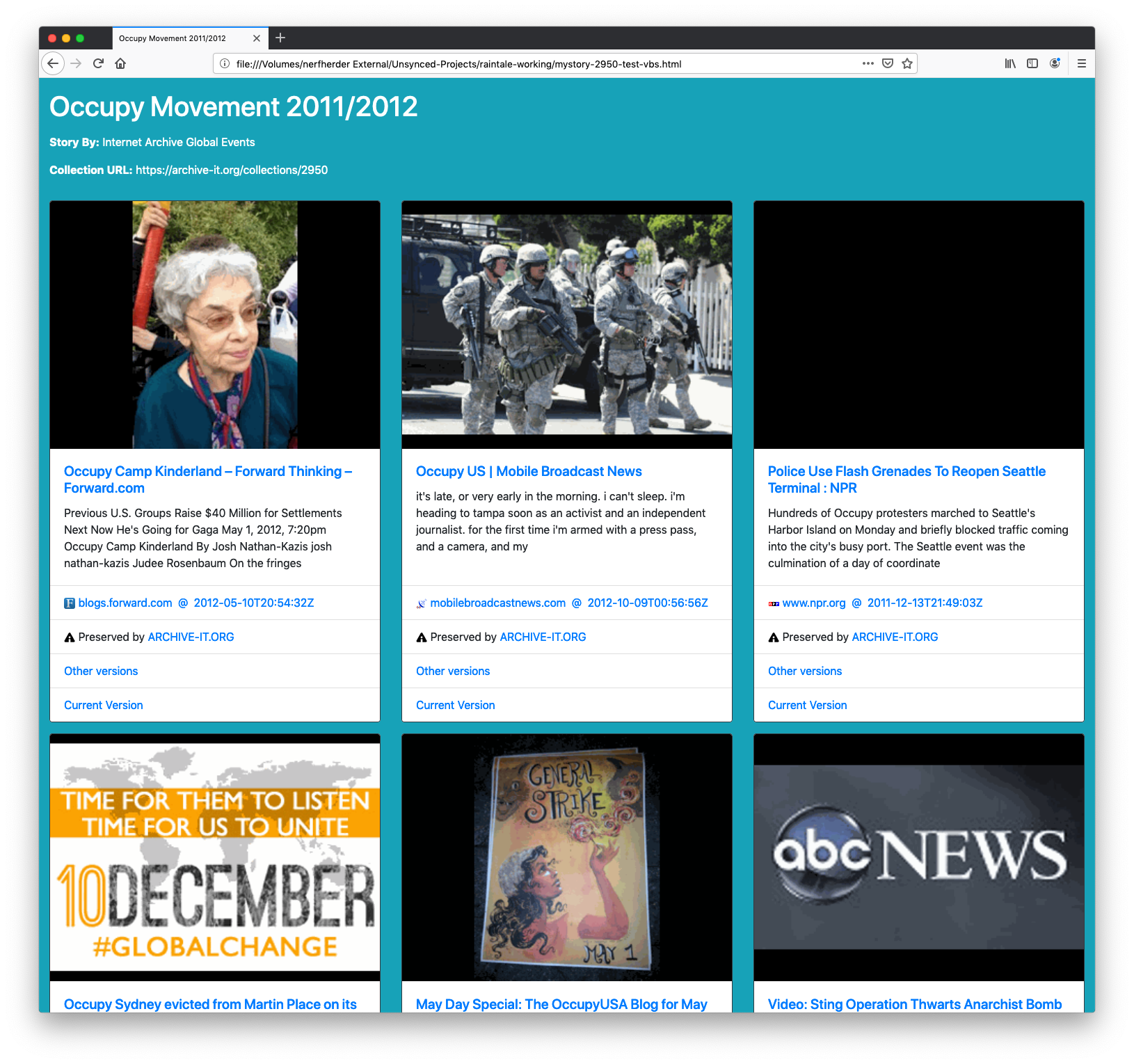}
  \caption{Raintale output examples - HTML with social cards created via Bootstrap}
  \label{fig:template-vbsir-html}
\end{figure}

\begin{figure}[H]

  \centering
  \includegraphics[height=\textheight]{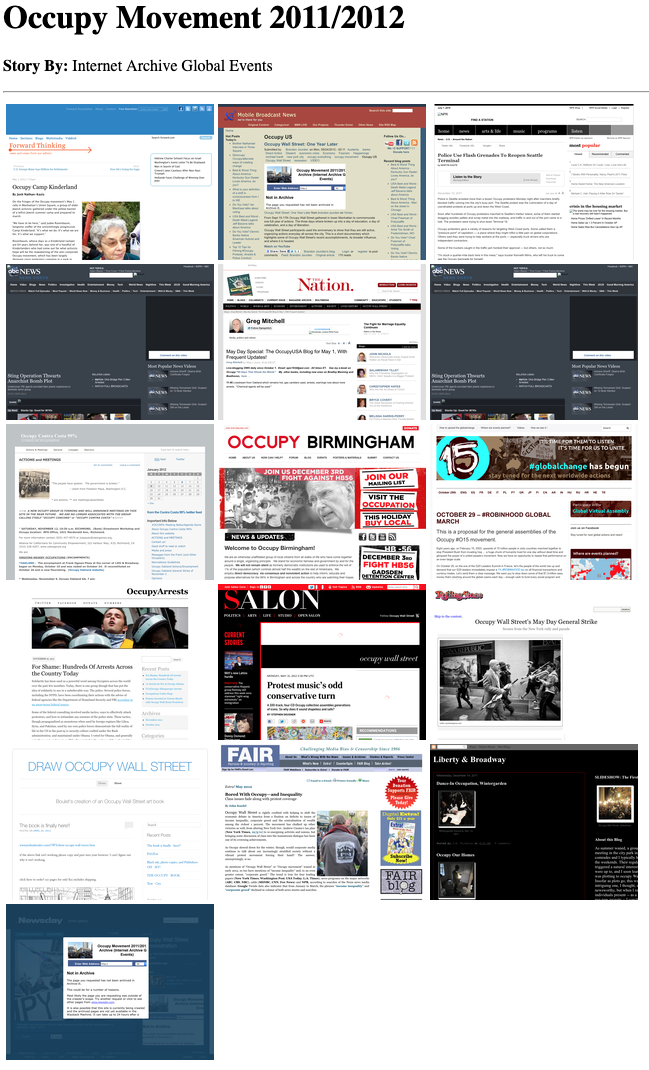}
  \caption{Raintale output examples - 3 column HTML story of thumbnails}
  \label{fig:template-thumbnails3col-html}

\end{figure}

\begin{figure}[H]

  \centering

  \includegraphics[width=\textwidth]{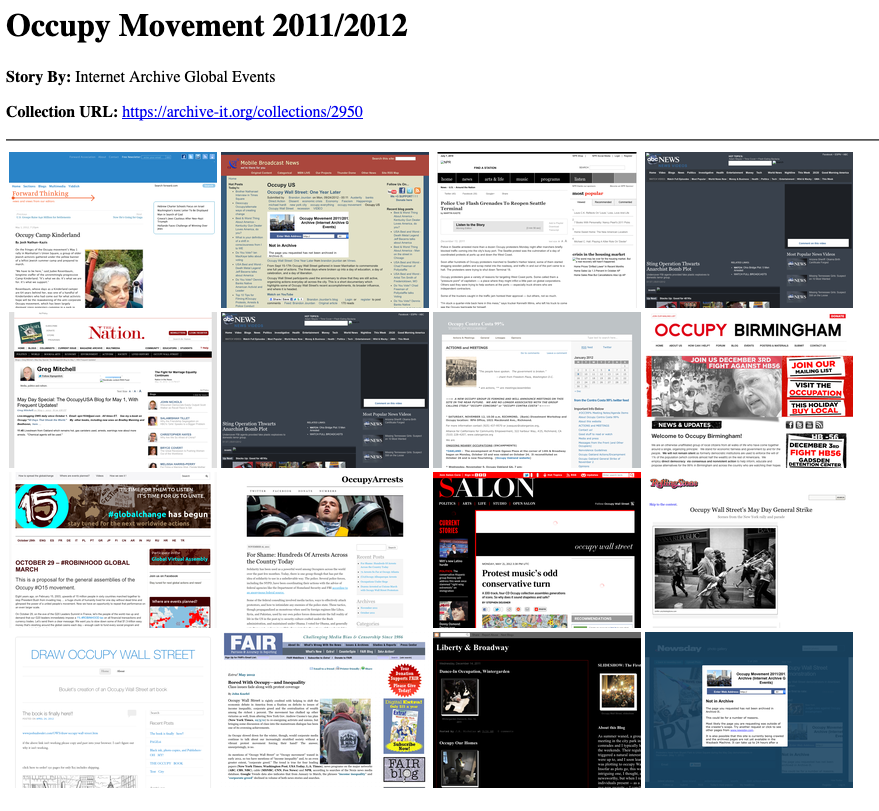}
  \caption{Raintale output examples - 4 column HTML story of thumbnails}
  \label{fig:template-thumbnails4col-html}
\end{figure}

\begin{figure}[H]

  \centering
  \includegraphics[height=\textheight]{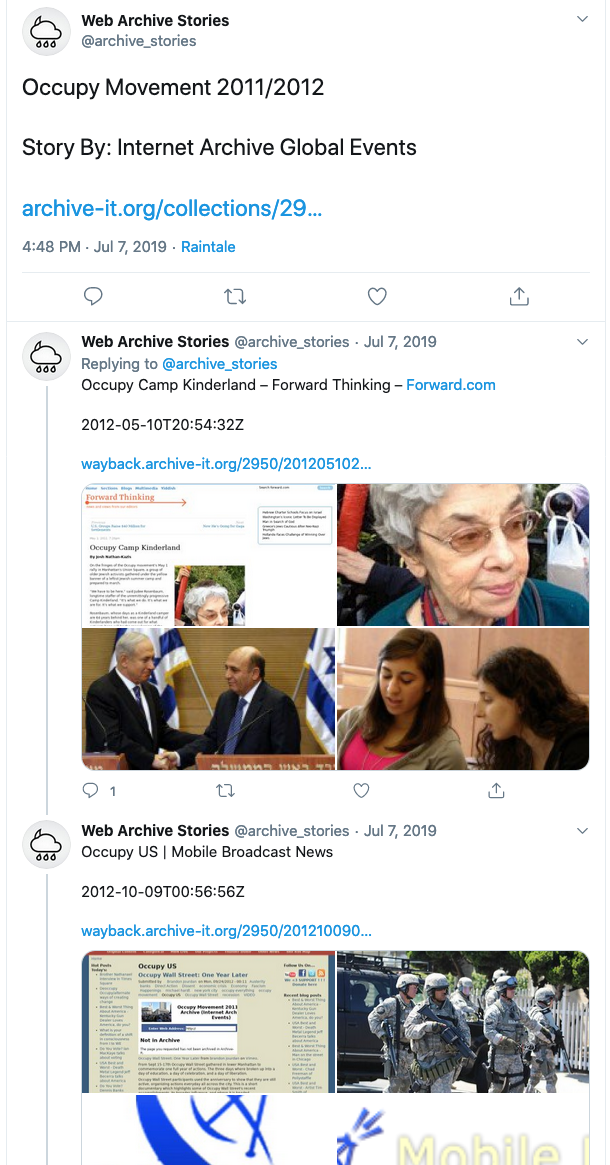}
  \caption{Raintale output examples - Twitter thread}
  \label{fig:twitter-example}
\end{figure}

\begin{figure}[H]

  \centering

  \includegraphics[height=\textheight]{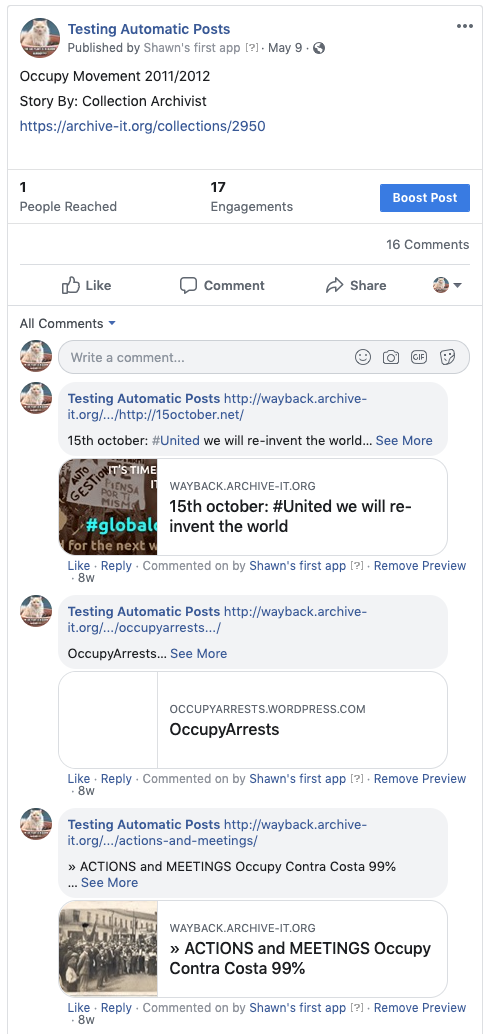}
  \caption{Raintale output examples - Facebook post}
  \label{fig:template-default-facebook}
\end{figure}

\begin{figure}[H]

  \centering
  \includegraphics[width=\textwidth]{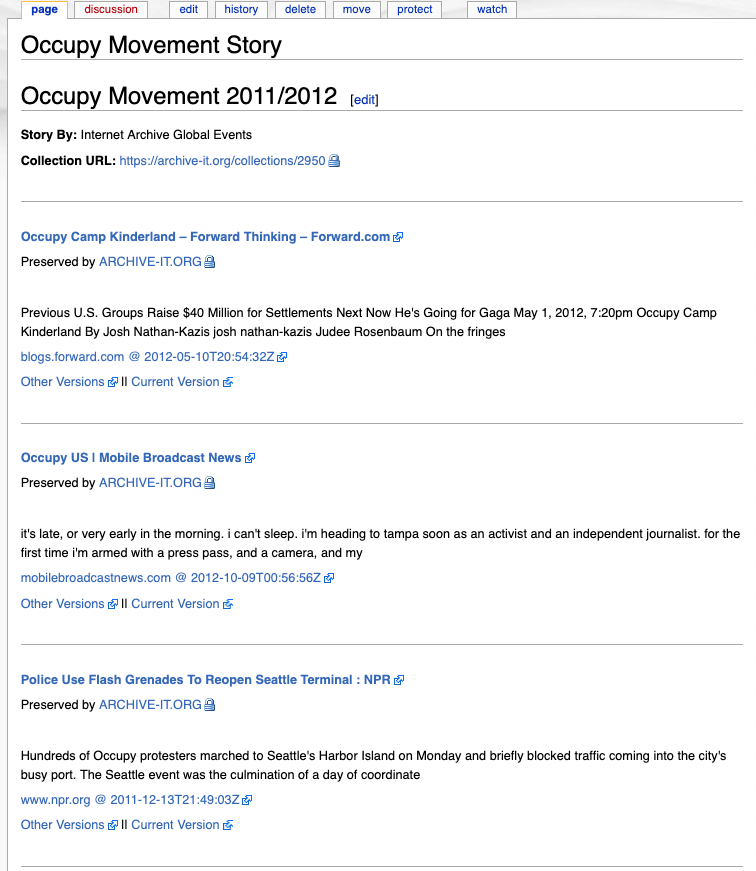}
  \caption{Raintale output examples - MediaWiki markup published to a MediaWiki article}
  \label{fig:template-default-mediawiki}
\end{figure}

\begin{figure}[H]

  \centering

  \includegraphics[width=\textwidth]{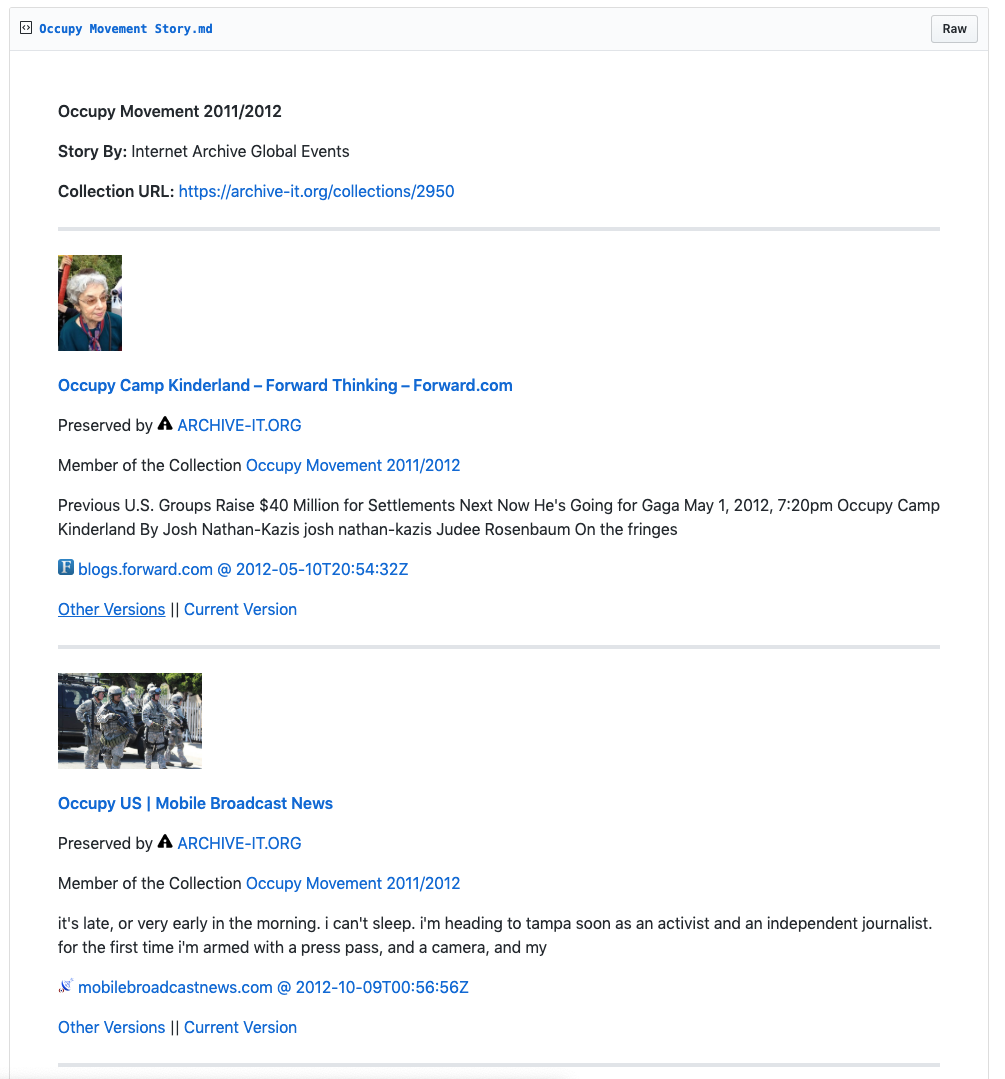}
  \caption{Raintale output examples - Markdown in a GitHub gist}
  \label{fig:template-default-markdown}
\end{figure}

Raintale supports different types of storytellers. File storytellers, such as \texttt{template} or \texttt{html} save content to a file. Service storytellers, such as \texttt{twitter} or \texttt{facebook}, write content to a specifc social media service. A user can apply templates to different types of storytellers. Instead of supplying their own template, a Raintale user can apply one of its pre-packaged \textbf{presets}. On pages \pageref{fig:raintale-template-default-html} to \pageref{fig:template-default-markdown}, Figures \ref{fig:raintale-template-default-html} to \ref{fig:template-default-markdown} display examples applying these presets for mementos from the Archive-It collection \emph{Occupy Movement 2011/2012}. Internally, a template exists for each of these and Raintale processes this pre-package template rather than requiring that the user supply their own.

For file storytellers, as demonstrated in Figures \ref{fig:raintale-template-default-html}, \ref{fig:template-vbsir-html}, \ref{fig:template-thumbnails3col-html}, \ref{fig:template-thumbnails4col-html}, \ref{fig:template-default-mediawiki}, and \ref{fig:template-default-markdown}, Raintale requires that the user supply the name of the output file. The following command generates a story with the preset HTML storyteller and the URI-Ms listed in the file \texttt{story-mementos.txt} and writes the output to a file named \texttt{mystory.html}.
\begin{lstlisting}[frame=none,postbreak=]
# tellstory -i story-mementos.txt --storyteller html -o mystory.html --title "This is My Story Title"
\end{lstlisting}

For service storytellers, such as Twitter (Figure \ref{fig:twitter-example}), the user must supply a YAML file containing the appropriate credentials to use that service. The command below generates a story with the preset Twitter storyteller and the URI-Ms listed in the file \texttt{story-mementos.txt} and writes the output as a Twitter thread using the credentials specified in \texttt{twitter-credentials.yml}.
\begin{lstlisting}[frame=none,postbreak=]
# tellstory -i story-mementos.txt --storyteller twitter --title "This is My Story Title" -c twitter-credentials.yml
\end{lstlisting}

% Raintale can produce many output formats, including HTML (Figure \ref{fig:html-example}), Markdown, MediaWiki, Jekyll, and Twitter Threads (Figure  \ref{fig:twitter-example}).

\begin{figure}[htbp]
\begin{lstlisting}
http://wayback.archive-it.org/2950/20120510205501/http://www.thenation.com/blog/167643/may-day-special-occupyusa-blog-may-1-frequent-updates/
http://wayback.archive-it.org/2950/20120814042704/http://occupyarrests.wordpress.com/
\end{lstlisting}
\caption{A text input file for Raintale} 
\label{fig:raintale-text-input}
\end{figure}

\begin{figure}[htbp]
  \centering
  \includegraphics[width=0.8\textwidth]{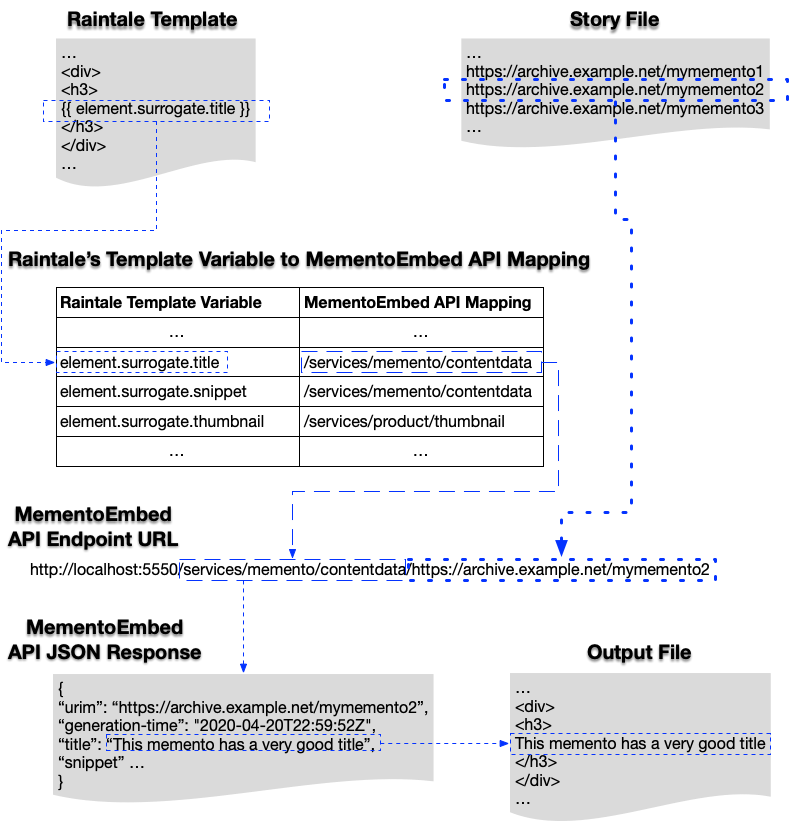}
  \caption{How Raintale combines the template and the story file to insert memento data into the desired output} 
  \label{fig:variable-to-api-mapping}
\end{figure}

\begin{figure}[htbp]
\begin{lstlisting}
{
    "title": "My Story Title",
    "collection_url": "https://archive.example.com/mycollection",
    "generated_by": "My Curator",
    "metadata": {
        "myKey1": "value1",
        "my-Key2": "value2",
        "my key 3": "value 3"
    },
    "elements": [
        {
            "type": "text",
            "value": "Livestream from Oakland which remains hot, gas canisters used, arrests, warnings now about more arrests.  \"Chemical agents will be used.\""
        },
        {
            "type": "link",
            "value": "http://wayback.archive-it.org/2950/20120510205501/http://www.thenation.com/blog/167643/may-day-special-occupyusa-blog-may-1-frequent-updates/"
        },
        {
            "type": "text",
            "value": "For Shame: Hundreds Of Arrests Across the Country Today"
        },
        {
            "type": "link",
            "value": "http://wayback.archive-it.org/2950/20120814042704/http://occupyarrests.wordpress.com/"
        }
    ]
}
\end{lstlisting}
\caption{A JSON input file for Raintale}
\label{fig:raintale-json-input}
\end{figure}

\begin{figure}[htbp]
  \begin{lstlisting}[language=html,numbers=left,stepnumber=1]
    <p><h1>{{ title }}</h1></p>
  
    {% if generated_by is not none %}
    <p><strong>Story By:</strong> {{ generated_by }}</p>
    {% endif %}
    
    {% if collection_url is not none %}
    <p><strong>Collection URI:</strong> <a href="{{ collection_url }}">{{ collection_url }}</a></p>
    {% endif %}
    
    <hr>
    
    <table border="0">
    <tr>
    {% for element in elements %}
    
    {% if element.type == 'link' %}
    
        <td><a href="{{ element.surrogate.urim }}"><img src="{{ element.surrogate.thumbnail|prefer remove_banner=yes }}"></a></td>
    
        {% if loop.index is divisibleby 4 %}
            </tr><tr>
        {% endif %}
    
    {% else %}
    
    <!-- Element type {{ element.type }} is unsupported by the thumbnails4col template -->
    
    {% endif %}
    
    {% endfor %}
    </table>
  \end{lstlisting}
  \caption{An example Raintale template for the story shown in Figure \ref{fig:template-thumbnails4col-html}}
  \label{fig:example-raintale-template}
\end{figure}

\begin{figure}[htbp]
  \begin{lstlisting}[language=html,numbers=left,stepnumber=1]
{# RAINTALE MULTIPART TEMPLATE #}
{# RAINTALE TITLE PART #}
{{ title }}

{% if generated_by is defined %}Story By: {{ generated_by }}{% endif %}

{% if collection_url is defined %}{{ collection_url }}{% endif %}
{# RAINTALE ELEMENT PART #}

{{ element.surrogate.title }}

{{ element.surrogate.memento_datetime }}

{{ element.surrogate.urim }}

{# RAINTALE ELEMENT MEDIA #}
{{ element.surrogate.thumbnail|prefer thumbnail_width=1024,remove_banner=yes }}
{{ element.surrogate.image|prefer rank=1 }}
{{ element.surrogate.image|prefer rank=2 }}
{{ element.surrogate.image|prefer rank=3 }}

\end{lstlisting}
\caption{An example Raintale template for the Twitter story shown in Figure \ref{fig:template-thumbnails4col-html}}
\label{fig:example-raintale-template-twitter}
\end{figure}

% \subsection{Raintale Inputs}

Raintale accepts two inputs: a story file containing URI-Ms and a template (or preset specification). The story file can take two forms, depending on the needs of the user.

The simplest story file format is a newline-separate text file containing URI-Ms. Figure \ref{fig:raintale-text-input} demonstrates an example. A Raintale user applying this format \emph{must} supply a title to their story via the \texttt{--title} command-line argument, as shown below.

\begin{lstlisting}[frame=none,postbreak=]
# tellstory -i story-mementos.txt --storyteller html -o mystory.html --title "This is My Story Title"
\end{lstlisting}

Where the story file contains the URI-Ms and the values of some of the template variables, the Raintale template controls the formatting and which information Raintale will request and display for each URI-M. Figure \ref{fig:variable-to-api-mapping} demonstrates a simplified example of how this works. Raintale first consumes the template file and determines which variables exist. At this point in the template, Raintale encounters HTML formatting and the Raintale variable \texttt{element.surrogate.title}. Raintale uses its internal mapping to discover that the appropriate MementoEmbed API endpoint for \texttt{element.surrogate.title} is \texttt{/services/memento/contentdata}. Then Raintale iterates through all elements in the story file. When it reaches \url{https://archive.example.net/mymemento2}, Raintale appends the current URI-M (\url{https://archive.example.net/mymemento2}) to \texttt{/services/memento/contentdata} and submits an HTTP request to MementoEmbed. It then extracts the title from MementoEmbed's JSON output and applies it to the resulting output file for the story.

Often users will want to specify more information for their stories than is possible with the newline-separated text file. For these use cases, they can supply a JSON formatted story file, as shown in Figure \ref{fig:raintale-json-input}. This JSON format allows the user to provide additional values that can be processed by the Raintale template engine. In Figure \ref{fig:raintale-json-input}, the \texttt{elements} key provides the list of items to include in the story. Each element of type \texttt{link} refers to a URI-M that Raintale will expand into a surrogate based on the template. Each element of type \texttt{text} contains a string that Raintale will insert into the story. Additionally, outside of the \texttt{elements} list, users can supply additional variables, such as \texttt{generated\_by}, that Raintale will apply to the corresponding variable in the template. In this case, Raintale users do not need to specify a value for the \texttt{--title} argument.

\begin{lstlisting}[frame=none,postbreak=]
  # tellstory -i story-data.json --storyteller html -o mystory.html
\end{lstlisting}

Figure \ref{fig:example-raintale-template} displays a more complex template that will produce the output shown in Figure \ref{fig:template-thumbnails4col-html}. On line 1, the \texttt{\{}\texttt{\{} \texttt{title} \texttt{\}}\texttt{\}} will be replaced with the title specified by the end user. The same will occur for the values of \texttt{generated\_by} and \texttt{collection\_url} on lines 4 and 8. On line 15, Raintale starts iterating through the list of story elements. On line 19, it will replace each \texttt{element.surrogate.urim} with the URI-M at that point in the iteration. On that same line, it will replace \texttt{element.surrogate.thumbnail} with a thumbnail generated by MementoEmbed for that same URI-M. The \texttt{prefer remove\_banner=yes} requests that the resulting thumbnail not contain a web archive's branding banner, a MementoEmbed preference. From this example we not only see how the variables are turned into formatted data, but also how a template writer can apply the MementoEmbed preferences listed in Table \ref{tab:preferences_api_services} of Appendix B to customize their story.

Raintale handles social media posts differently than file outputs. It assumes that social media stories will take the form of a single post, followed by comments. Figure \ref{fig:example-raintale-template-twitter} contains a Raintale \textbf{multipart template} used to generate the Tweets shown in Figure \ref{fig:twitter-example}. This template consists of three parts. The \texttt{RAINTALE TITLE PART} contains the variables and formatting of the first post in the Twitter thread. In this case, the first post consists of the story title, who or what generated the selection of mementos for this story, and the URI of the collection containing these mementos. The \texttt{RAINTALE ELEMENT PART} contains the variables and formatting of the response posts of the Twitter thread, here the title, memento-datetime, and URI-M of each memento. Twitter will turn the URI-M into a clickable link. Finally, the \texttt{RAINTALE ELEMENT MEDIA} section contains the media that Raintale will include in each response post. This last section will instruct Raintale to upload the thumbnail generated by MementoEmbed, and the top three scoring images from the memento. When using such templates for social media, we must consider the limitations of various social media services, such as Twitter's restrictions of four images per tweet and 280 characters of text. Raintale may partially post a story to Twitter if one of the tweets in the story fails to account for one of these restrictions.

Table \ref{tab:raintale_variables} of Appendix B contains a list of variables that Raintale users can include in their own templates. Through these variables and their own formatting, Raintale template authors can effectively generate their own surrogates. Recall from Section \ref{sec:card_services} that memento surrogates need special consideration to avoid misleading users. Also, while social media services require specific storytellers, Raintale's file output formats are not limited to HTML or Markdown. Any text-based singular file format can be supported, like JSON, XML, or CSV, making it quite valuable for a variety of automated workflows.

Raintale also supports an experimental video storytelling output inspired by the MementoEmbed docreel. Where the docreel represents a single document, this video fades in sentences and images from multiple documents. When deciding which content to include in the video, Raintale leverages the scores of the images and sentences provided by MementoEmbed. A user can create one of these videos when they wish to feature a small sample of a collection in a single social media post.

\section{Future Work}

Many potential directions exist for these tools. We can expand MementoEmbed's API to include summarization by more algorithms than Readability, JusText, Lede3, and TextRank. We may apply the sumy Python library \cite{sumy_python} to introduce other automatic extractive summarization algorithms like LexRank \cite{erkan2004lexrank}, Luhn \cite{luhn_automatic_1958}, Edmundson \cite{10.1145/321510.321519}, Latent Semantic Analysis \cite{steinberger2004using}, SumBasic \cite{vanderwende2007beyond}, or Kullback-Lieber (KL) Sum \cite{haghighi-vanderwende-2009-exploring}. To assist with this text analysis, we might also allow API clients to specify their desired boilerplate removal method. We are currently exploring better image scoring and selection techniques than Equation \ref{eq:mementoembed_scoring}.

Raintale has the potential to be a story development tool that can support multiple external APIs. We can add new Raintale variables for querying CarbonDate \cite{10.1145/2487788.2488121}, Memento Damage \cite{brunelle_not_2015}, and other web archive APIs. Raintale is currently a command-line application but would benefit from a GUI for easier use. Finally, we intend to include additional output options. We have explored, but not yet implemented, direct publishing to Blogger, GitHub, MediaWiki, Instagram, Tumblr, and Pinterest. Raintale could also conceivably create output in PowerPoint or PDF for sharing memento information in places outside of the Web.

\section{Conclusions}

We introduced MementoEmbed for generating surrogates for single mementos and Raintale for generating complete stories of memento sets. We envision Raintale and MementoEmbed to be critical components for summarizing collections of archived web pages through visualizations familiar to general users. We developed MementoEmbed so that its API is easily usable by machine clients. Raintale is a client of MementoEmbed that allows archivists to generate complete stories for lists of mementos. As a command-line application, archivists can incorporate Raintale into existing automated archiving workflows. Archivists can leverage this form of storytelling to highlight a specific subset of mementos from a collection. They can advertise their holdings, feature specific perspectives, focus on individual mementos, or help users decide if a collection meets their needs.

\bibliographystyle{acm}
\bibliography{references}

\clearpage
\section*{Appendix A: MementoEmbed Detail Tables}

\begin{table}[htbp]
  \caption{The fields provided by a successful response for the API endpoints under \texttt{/services/memento/}.}
  \label{tab:services_memento_fields}
  \begin{tabular}{l|l|p{2.5in}}
  \textbf{API Endpoint}                                    & \textbf{Field}              & \textbf{Field Description}                                                                                                                                \\ \hline
  /services/memento/contentdata/          & title                       & The title discovered in the memento                                                                                                                       \\ \cline{2-3} 
                                                           & snippet                     & The snippet discovered or generated from the memento                                                                                                      \\ \hline
  /services/memento/bestimage/                             & best-image-uri              & The striking image discovered in or generated from the memento                                                                                            \\ \hline
  /services/memento/imagedata/            & images                      & a JSON object containing each image URI-M and its calculated attributes (e.g., pixel size)                                                                \\ \cline{2-3} 
                                                           & ranked images               & a JSON object listing each image URI-M in descending order by score                                                                                       \\ \hline
  /services/memento/archivedata/          & archive-uri                 & The URI of the archive's home page                                                                                                                        \\ \cline{2-3} 
                                                           & archive-name                & The upper case domain name of the archive                                                                                                                 \\ \cline{2-3} 
                                                           & archive-favicon             & The detected favicon URI for the archvie                                                                                                                  \\ \cline{2-3} 
                                                           & archive-collection-id       & \begin{tabular}[c]{@{}l@{}}The Collection ID for the URI-M;\\ only for Archive-It mementos\end{tabular}                                                   \\ \cline{2-3} 
                                                           & archive-collection-name     & \begin{tabular}[c]{@{}l@{}}The Collection Name for the URI-M;\\ only for Archive-It mementos\end{tabular}                                                 \\ \cline{2-3} 
                                                           & archive-collection-uri      & \begin{tabular}[c]{@{}l@{}}The URI of the collection;\\ only for Archive-It mementos\end{tabular}                                                         \\ \hline
  /services/memento/originalresourcedata/ & original-uri                & The URI-R of the memento                                                                                                                                  \\ \cline{2-3} 
                                                           & original-domain             & The domain of the URI-R                                                                                                                                   \\ \cline{2-3} 
                                                           & original-favicon            & The favicon discovered for this memento                                                                                                                   \\ \cline{2-3} 
                                                           & original-linkstatus         & Is ``Live'' if the URI-R still responds with an HTTP 200 status code                                                                                        \\ \hline
  /services/memento/seeddata/             & timemap                     & The URI-T of the memento                                                                                                                                  \\ \cline{2-3} 
                                                           & original-url                & The URI-R of the memento                                                                                                                                  \\ \cline{2-3} 
                                                           & memento-count               & The number of mementos in this memento's TimeMap                                                                                                          \\ \cline{2-3} 
                                                           & first-memento-datetime      & The memento-datetime of the first memento in the TimeMap                                                                                                  \\ \cline{2-3} 
                                                           & first-urim                  & The first URI-M in this memento's TimeMap                                                                                                                 \\ \cline{2-3} 
                                                           & last-memento-datetime       & The memento-datetime of the last memento in the TimeMap                                                                                                   \\ \cline{2-3} 
                                                           & last-urim                   & The last URI-M in this memento's TimeMap                                                                                                                  \\ \cline{2-3} 
                                                           & metadata                    & \begin{tabular}[c]{@{}l@{}}Metadata associated with the URI-R of this \\ memento, if it was a seed;\\ only available for Archive-It collections\end{tabular} \\ \hline
  /services/memento/paragraphrank/        & algorithm                   & The algorithm used to score the memento's paragraphs                                                                                                      \\ \cline{2-3} 
                                                           & scored paragraphs           & Each paragraph extracted from the memento and its score                                                                                                   \\ \hline
  /services/memento/sentencerank/         & paragraph scoring algorithm & The algorithm used to score the memento's paragraphs                                                                                                      \\ \cline{2-3} 
                                                           & sentence ranking algorithm  & The algorithm used to score the memento's sentences within each paragraph                                                                                 \\ \hline
  /services/memento/page-metadata/                         & page-metadata               & A JSON object containing the values extracted from the memento's META elements                                                                            \\ 

  \end{tabular}
\end{table}

\begin{table}[t]
  \caption{The preferences available for various MementoEmbed API endpoints.}
  \label{tab:preferences_api_services}
  \tiny
  \begin{tabular}{l|l|p{1.6in}|p{0.7in}|p{0.7in}}
  \textbf{API Endpoint}           & \textbf{Preference Name}       & \textbf{Preference Description}                                                                                           & \textbf{Possible Values}                                   & \textbf{Default Value} \\ \hline
  /services/memento/sentencerank/ & algorithm                 & The paragraph and sentence scoring algorithms to use                                                                      & readability/lede3, readability/textrank, justext/textrank  & readability/lede3      \\ \hline
  /services/product/socialcard/   & datauri\_favicon          & 'yes' instructs MementoEmbed to generate data URIs for favicons                                                           & yes, no                                                    & no                     \\ \cline{2-5}
                                  & datauri\_image            & 'yes' instructs MementoEmbed to generate data URIs for striking images                                                 & yes, no                                                    & no                     \\ \cline{2-5}
                                  & using\_remote\_javascript & 'no' instructs MementoEmbed to construct HTML that relies on no external JavaScript                                       & yes, no                                                    & yes                    \\ \cline{2-5}
                                  & minify\_markup            & 'yes' instructs MementoEmbed to minify the output                                                                         & yes, no                                                    & no                     \\ \hline
  /services/product/thumbnail/    & viewport\_width           & the width of the viewport in pixels for the browser capturing the screenshot                                              & \textless{}= 5120                                          & 1024 px                \\ \hline
                                  & viewport\_height          & the height of the viewport in pixels for the browser capturing the screenshot                                             & \textless{}= 2880                                          & 768 px                 \\ \cline{2-5}
                                  & thumbnail\_width          & the width of the thumbnail in pixels for the browser capturing the screenshot                                             & \textless{}= 5120                                          & 208 px                 \\ \cline{2-5}
                                  & thumbnail\_height         & the height of the thumbnail in pixels for the browser capturing the screenshot                                            & \textless{}= 2880                                          & 156 px                 \\ \cline{2-5}
                                  & timeout                   & how long MementoEmbed should wait for Puppeteer to create a thumbnail                                                     & \textless{}= 300                                           & 300 seconds            \\ \cline{2-5}
                                  & remove\_banner            & 'yes' instructs MementoEmbed to try to remove the archive-specific banner from the memento prior to taking the screenshot & yes, no                                                    & no                     \\ \hline
  /services/product/imagereel/    & duration                  & the length, in seconds, between image transitions, including fades                                                        & \textless{}= 300                                           & 100 seconds            \\ \hline
                                  & imagecount                & the maximum number of images to include                                                                                   & \textless{}= 10                                            & 5                      \\ \cline{2-5}
                                  & width                     & the width of the imagereel in pixels                                                                                      & \textless{}= 5120                                          & 320 px                 \\ \cline{2-5}
                                  & height                    & the height of the imagereel in pixels                                                                                     & \textless{}= 2880                                          & 240 px                 \\ \cline{2-5}
  /services/product/wordcloud/    & colormap                  & the matplotlib colormap of the Word Cloud                                                                                 & See \cite{matplotlib_colormaps} for a list of values & inferno                \\ \cline{2-5}
                                  & background\_color         & the background color of the word cloud                                                                                    & hexadecimal values, RGB, HSL, HSV, common HTML color names & white                  \\ \cline{2-5}
                                  & textonly                  & 'yes' instructs MementoEmbed to return a JSON list of words instead of a word cloud                                       & yes, no                                                    & no                     \\ \hline
  /services/product/docreel/      & duration                  & the duration between transitions, including fades                                                                         & \textless{}= 300                                           & 100                    \\ \hline
                                  & imagecount                & the number of images to include                                                                                           & \textless{}= 10                                            & 5                      \\ \cline{2-5}
                                  & sentencecount             & the number of sentences to include                                                                                        & \textless{}= 10                                            & 5                      \\ \cline{2-5}
                                  & width                     & the width of the docreel in pixels                                                                                        & \textless{}= 5120                                          & 320 px                 \\ \cline{2-5}
                                  & height                    & the height of the docreel in pixels                                                                                       & \textless{}= 2880                                          & 240 px                 \\
  \end{tabular}
\end{table}

\clearpage
\section*{Appendix B: Raintale Detail Tables}

\begin{table}[t]
  \caption{Raintale variables available for individual mementos in a template}
  \label{tab:raintale_variables}
  \begin{tabular}{l|p{3.8in}}
  \textbf{Raintale template variable}         & \textbf{Description}                                                                                                                                                                                                    \\ \hline
  \texttt{element.surrogate.archive\_collection\_id}   & the ID of the collection containing this URI-M \\
  & only works with URI-Ms from public Archive-It collections \\ \hline
  \texttt{element.surrogate.archive\_collection\_name} & the name of the collection containing this URI-M \\
  & only works with URI-Ms from public Archive-It collections \\ \hline
  \texttt{element.surrogate.archive\_collection\_uri}  & the URI of the collection containing this URI-M \\
  & only works with URI-Ms from public Archive-It collections \\ \hline
  \texttt{element.surrogate.archive\_favicon}         & the URI of the favicon of the archive containing this URI-M \\ \hline
  \texttt{element.surrogate.archive\_name}            & the uppercase name of the domain name of the archive containing this URI-M \\ \hline
  \texttt{element.surrogate.archive\_uri}             & the URI of the archive containing this URI-M \\ \hline
  \texttt{element.surrogate.best\_image\_uri}         & the URI of the best image found in the memento using the MementoEmbed image scoring equation; will be removed in the future in favor of \\
  & \texttt{element.surrogate.image|prefer rank=1} \\ \hline
  \texttt{element.surrogate.first\_memento\_datetime} & the datetime of the earliest memento for this resource at the web archive containing this URI-M \\ \hline
  \texttt{element.surrogate.first\_urim}              & the URI-M of the earliest memento for this resource at the web archive containing this URI-M \\ \hline
  \texttt{element.surrogate.image}                    & provides the i\textsuperscript{th} best image found in the memento using the MementoEmbed image scoring equation;this allows a user to include multiple images from a memento in a surrogate \\ \hline
  \texttt{element.surrogate.imagereel}                    & provides an imagreel for the memento, as generated by MementoEmbed \\ \hline
  \texttt{element.surrogate.last\_memento\_datetime}  & the datetime of the latest memento for this resource at the web archive containing this URI-M \\ \hline
  \texttt{element.surrogate.last\_urim}               & the URI-M of the latest memento for this resource at the web archive containing this URI-M \\ \hline
  \texttt{element.surrogate.memento\_count}           & the number of other mementos for this same resource at the web archive containing this URI-M \\ \hline
  \texttt{element.surrogate.memento\_datetime}        & the memento-datetime of the memento \\ \hline
  \texttt{element.surrogate.memento\_datetime\_14num} & provides the memento-datetime in YYYYMMDDHHMMSS format, a special datetime format for use in some web archive URI-Ms and API calls; will be removed in the future in favor of \\
  & \verb+element.surrogate.memento_datetime.strftime('\%Y\%m\%d\%H\%M\%S')+ \\ \hline
  \texttt{element.surrogate.metadata}                 & the metadata provided for the seed of this URI-M; provides a JSON object of keys and values corresponding   to the metadata fields present at the archive - only works with URI-Ms from   public Archive-It collections \\ \hline
  \texttt{element.surrogate.original\_domain}          & the domain of the original resource for this URI-M \\ \hline
  \texttt{element.surrogate.original\_favicon}         & the favicon corresponding to the original resource for this URI-M \\ \hline
  \texttt{element.surrogate.original\_linkstatus}      & the status of the link at the time the story was generated \\ \hline
  \texttt{element.surrogate.original\_uri}             & the URI of the original resource (URI-R) corresponding to this memento \\ \hline
  \texttt{element.surrogate.sentence}                  & the i\textsuperscript{th} best sentence found in the memento using the MementoEmbed sentence scoring equation; this allows a user to include multiple sentences from a memento in a surrogate \\ \hline
  \texttt{element.surrogate.snippet}                   & the best text extracted from the memento via MementoEmbed \\ \hline
  \texttt{element.thumbnail}                           & provides a thumbnail generated by MementoEmbed for the given URI-M as a data URI \\ \hline
  \texttt{element.surrogate.timegate\_uri}             & the URI of the Memento TimeGate (URI-G) corresponding to the resource \\ \hline
  \texttt{element.surrogate.timemap\_uri}              & the URI of the Memento TimeMap (URI-T) corresponding to the resource \\ \hline
  \texttt{element.surrogate.title}                     & the text of the title extracted from the HTML of the memento \\ \hline
  \texttt{element.surrogate.urim}                      & the URI-M of this memento \\
  \end{tabular}
\end{table}

\begin{table}[t]
  \begin{tabular}{l|l}
  \textbf{Raintale template variable}         & \textbf{Corresponding MementoEmbed API Endpoint}                 \\ \hline
  \texttt{element.surrogate.archive\_collection\_id}   & /services/memento/archivedata/                                   \\ \hline
  \texttt{element.surrogate.archive\_collection\_name} & /services/memento/archivedata/                                   \\ \hline
  \texttt{element.surrogate.archive\_collection\_uri}  & /services/memento/archivedata/                                   \\ \hline
  \texttt{element.surrogate.archive\_favicon}          & /services/memento/archivedata/                                   \\ \hline
  \texttt{element.surrogate.archive\_name}             & /services/memento/archivedata/                                   \\ \hline
  \texttt{element.surrogate.archive\_uri}              & /services/memento/archivedata/                                   \\ \hline
  \texttt{element.surrogate.best\_image\_uri}          & /services/memento/bestimage/                                     \\ \hline
  \texttt{element.surrogate.first\_memento\_datetime}  & /services/memento/seeddata/                                      \\ \hline
  \texttt{element.surrogate.first\_urim}               & /services/memento/seeddata/                                      \\ \hline
  \texttt{element.surrogate.image}                     & /services/memento/imagedata/                                     \\ \hline
  \texttt{element.surrogate.imagereel}                 & /services/product/imagereel/                                     \\ \hline
  \texttt{element.surrogate.last\_memento\_datetime}   & /services/memento/seeddata/                                      \\ \hline
  \texttt{element.surrogate.last\_urim}                & /services/memento/seeddata/                                      \\ \hline
  \texttt{element.surrogate.memento\_count}            & /services/memento/seeddata/                                      \\ \hline
  \texttt{element.surrogate.memento\_datetime}         & /services/memento/contentdata/                                   \\ \hline
  \texttt{element.surrogate.memento\_datetime\_14num}  & Calculated by Raintale based on   /services/memento/contentdata/ \\ \hline
  \texttt{element.surrogate.metadata}                  & /services/memento/seeddata/                                      \\ \hline
  \texttt{element.surrogate.original\_domain}          & /services/memento/originalresourcedata/                          \\ \hline
  \texttt{element.surrogate.original\_favicon}         & /services/memento/originalresourcedata/                          \\ \hline
  \texttt{element.surrogate.original\_linkstatus}      & /services/memento/originalresourcedata/                          \\ \hline
  \texttt{element.surrogate.original\_uri}             & /services/memento/originalresourcedata/                          \\ \hline
  \texttt{element.surrogate.sentence}                  & /services/memento/sentencerank/                                  \\ \hline
  \texttt{element.surrogate.snippet}                   & /services/memento/contentdata/                                   \\ \hline
  \texttt{element.thumbnail}                           & /services/product/thumbnail/                                     \\ \hline
  \texttt{element.surrogate.timegate\_uri}             & /services/memento/seeddata/                                      \\ \hline
  \texttt{element.surrogate.timemap\_uri}              & /services/memento/seeddata/                                      \\ \hline
  \texttt{element.surrogate.title}                     & /services/memento/contentdata/                                   \\ \hline
  \texttt{element.surrogate.urim}                      & N/A - based on input from story   file                           \\
  \end{tabular}
\end{table}

\end{document}